\documentclass[fleqn,usenatbib]{mnras}

\usepackage{newtxtext,newtxmath}
\usepackage{float}
\usepackage{array}
\usepackage{makecell}
\usepackage{longtable}
\usepackage{hyperref}
\usepackage{anyfontsize}
\usepackage[labelfont=bf,labelsep=period]{caption}
\usepackage[T1]{fontenc}
\usepackage{enumitem}

\DeclareRobustCommand{\VAN}[3]{#2}
\let\VANthebibliography\thebibliography
\def\thebibliography{\DeclareRobustCommand{\VAN}[3]{##3}\VANthebibliography}

\usepackage{graphicx}
\usepackage{amsmath}

\def\HI{H\,{\sc i}}
\def\3DB{$^\mathrm{3D}$\textsc{barolo}}

\title[Multi-resolution modelling of MHONGOOSE galaxies]{Multi-resolution kinematic modelling of nearby galaxies: a demonstration using MHONGOOSE observations}
\author[B. R. Makinson et al.]{
Bradley R. Makinson,$^{1,2}$\thanks{E-mail: bradley.r.makinson@gmail.com}
Kyle A. Oman,$^{1,2}$\thanks{E-mail: kyle.a.oman@durham.ac.uk}
Mark Swinbank$^{1}$
\\
$^{1}$Centre for Extragalactic Astronomy, Department of Physics, Durham University, South Road, Durham, DH1 3LE, UK\\
$^{2}$Institute for Computational Cosmology, Department of Physics, Durham University, South Road, Durham, DH1 3LE, UK\\
}

\date{Accepted XXX. Received YYY; in original form ZZZ}

\pubyear{\the\year{}}

\begin{document}
\label{firstpage}
\pagerange{\pageref{firstpage}--\pageref{lastpage}}
\maketitle

\begin{abstract}
We present a novel method of combining kinematic models obtained at multiple spatial resolution levels in a self-consistent manner. The MHONGOOSE survey has mapped atomic hydrogen emission in $30$ nearby dwarf and spiral galaxies. Each galaxy is imaged at multiple resolution levels with unprecedented dynamic range in spatial resolution (from $\sim 10''$ to $ 90''$) and \HI{} sensitivity, with the latter varying by almost a factor of $30$ across all resolution scales. We use radial weighting functions to combine kinematic models from all resolution levels. The weights are derived from the residuals of model fits to a set of observations of synthetic model galaxies with known rotation curves and geometries. We obtain combined (weighted and smoothed) inclination and position angle profiles for each galaxy. These suppress the sharp, often unphysical radial fluctuations arising in single-resolution profiles. We then fit the rotation speed and velocity dispersion profiles at each resolution level with the geometric profiles fixed to the combined profiles, finally combining these using the same weighting and smoothing approach. The combined rotation curves utilise all of the available information and have smaller typical systematic errors compared to those obtained using a single resolution level, particularly near the centres and outer edges of models. This initial demonstration is promising; there is scope to further refine the process to use such information-rich observations to their full potential.

\end{abstract}


\section{Introduction}

Rotation curves allow us to infer the dynamical mass distribution of galaxies and test the validity of cosmological models. Since the early $20^\mathrm{th}$ century, the rotation speeds of galaxies have been measured using the Doppler shift of optical emission lines, with H$\alpha$ emission serving as a primary tracer of ionized hydrogen in star-forming regions. However, this method was hampered by the limited radial extent of detectable H$\alpha$ emission in galaxies. The distribution of atomic hydrogen (\HI) detectable in $21\,\mathrm{cm}$ emission is often far more extended than that of its ionised counterpart. Radio observations enable rotation curves to be extended to larger radii, revealing that they remain flat well beyond the optical disk. This provided some of the earliest compelling evidence for the presence of dark matter \citep[e.g.][]{Rogstad1972,Roberts1973,Bosma1978}. Whilst early radio observations used single-dish telescopes, radio interferometers enable reaching much higher spatial resolutions and remain one of the most powerful tools for probing galaxy kinematics today. 

More recently, interest has focused on the rotation curves of dwarf galaxies, driven primarily by efforts to address long-standing issues such as the core-cusp problem and the origin of the diversity in rotation curve shapes of dwarf galaxies \citep[e.g.][see also \citealp{Sales2022} for a review]{Moore1994,Flores1994,deBlok2010,Oman2015,Santos2020}, highlighting the importance of rotation curves in a cosmological context. Since many of these galaxies are small and contain little baryonic mass, kiloparsec-scale resolution and high \HI{} column density sensitivity ($\lesssim 10^{19}\,\mathrm{cm}^{-2}$) are required to obtain detailed kinematics whilst maximising radial extent. Interferometric surveys such as HALOGAS\footnote{Hydrogen Accretion in LOcal GAlaxieS} \citep{HALOGAS} and LITTLE THINGS\footnote{Local Irregulars That Trace Lumiosity Extremes: The \HI{} Nearby Galaxies Survey} \citep{LITTLETHINGS} provide either high resolution or high sensitivity, but are suboptimal at providing both simultaneously.

The MHONGOOSE\footnote{MeerKAT \HI{} Observations of Nearby Galactic Objects: Observing Southern Emitters} survey breaks this trend. It uses the MeerKAT radio telescope array in South Africa to acquire deep \HI{} observations of $30$ nearby late type galaxies. MeerKAT is an array of $64$ $13.5\,\mathrm{m}$ dishes with a maximum baseline of $7.7\,\mathrm{km}$, with 70~per~cent of the collecting area located very close to the core of the array. This provides an unmatched combination of spatial resolution and column density sensitivity \citep[see][fig.~1]{MHONGOOSE}, offering the opportunity to study galaxies out to larger radii with finer detail than previously achievable.

\begin{table*}
  \centering
  \caption{Properties of the galaxies investigated in this work. All values are extracted from \citet{MHONGOOSE} unless otherwise stated. (1) HIPASS identification. (2) Alternative galaxy name. (3) Right ascension. (4) Declination. (5) Distance from Earth. (6) Galaxy inclination relative to line of sight. (7) Base-10 logarithm of the total \HI{} mass. (8) Base-10 logarithm of the total stellar mass. (9) Morphological types from HyperLeda \citep{HyperLEDA}.}
  \label{table: Targets}
  \begin{tabular}{ >{\centering\arraybackslash}m{1.7cm}  >{\centering\arraybackslash}m{2cm}  >
  {\centering\arraybackslash}m{2cm} >
  {\centering\arraybackslash}m{2cm} >
  {\centering\arraybackslash}m{1cm} >
  {\centering\arraybackslash}m{1cm} >
  {\centering\arraybackslash}m{1.5cm} >
  {\centering\arraybackslash}m{1.5cm} >
  {\centering\arraybackslash}m{1.2cm}}
  \hline
    \makecell{HIPASS \\ \\ (1)} &
    \makecell{Name \\ \\ (2)} &
    \makecell{$\alpha$ (J2000) \\ ($^h\ ^m\ ^s$) \\ (3)} &
    \makecell{$\delta$ (J2000) \\ ($^\circ\, '\, ''$) \\ (4)} &
    \makecell{$D$ \\ (Mpc) \\ (5)} &
    \makecell{$i$ \\ ($^\circ$) \\ (6)} &
    \makecell{$\log(M_{\text{HI}})$ \\ ($M_{\odot}$) \\ (7)} &
    \makecell{$\log(M_{\ast})$ \\ ($M_{\odot}$) \\ (8)} &
    \makecell{Morph. \\ \\ (9)}
    \\
    \hline
    
    J1106-14 & KKS2000-23 & 11 06 12.0 & –14 24 25.7 & 13.9 & 70 & 8.74 & 7.51 & Ir \\
    J0309-41 & ESO300-G014 & 03 09 37.9 & –41 01 49.7 & 10.9 & 59 & 8.89 & 8.90 & SABm \\

    \hline
  \end{tabular}
  \vspace{-0pt}
\end{table*}

\begin{table*}
\centering
\caption{Properties of MHONGOOSE data used for J1106-14 and J0309-41, at 5 of the standard resolution levels provided by the survey. (1) Resolution level label. (2) Spatial pixel size. (3) Average major-axis beam size. (4) Average minor-axis beam size. (5) Average position angle of the major axis of the beam. (6) Average noise per velocity channel. (7) 1$\sigma$, one-channel column density sensitivity. (8) 3$\sigma$ column density sensitivity integrated over a 16 km s$^{-1}$ velocity channel.}
\label{table: resolution levels}
\setlength{\tabcolsep}{10pt}
\begin{tabular}{lccccccc}
\hline
Label & Pixel size & $b_{\mathrm{maj}}$ & $b_{\mathrm{min}}$ & $b_{\mathrm{PA}}$ & Noise & $\log N_{\mathrm{HI}}^{1\sigma,\mathrm{1ch}}$ & $\log N_{\mathrm{HI}}^{3\sigma,16\mathrm{km~s}^{-1}}$ \\
 & (arcsec) & (arcsec) & (arcsec) & ($^\circ$) & (mJy beam$^{-1}$) & (cm$^{-2}$) & (cm$^{-2}$) \\
\hline
\multicolumn{7}{l}{\textit{J1106--14}} \\
r10\_t90 & 30.0 & 94.4 & 91.1 & 38.5  & 0.318 $\pm$ 0.011 & 16.76 & 17.77 \\
r05\_t60 & 20.0 & 64.8 & 64.1 & 94.8  & 0.250 $\pm$ 0.009 & 16.97 & 17.98 \\
r15\_t0  & 7.0  & 33.8 & 24.6 & 144.0 & 0.154 $\pm$ 0.004 & 17.44 & 18.44 \\
r10\_t0  & 5.0  & 26.2 & 17.6 & 143.2 & 0.150 $\pm$ 0.004 & 17.69 & 18.69 \\
r05\_t0  & 3.0  & 14.4 & 9.4  & 140.2 & 0.171 $\pm$ 0.005 & 18.29 & 19.29 \\
\hline
\multicolumn{7}{l}{\textit{J0309--41}} \\
r10\_t90 & 30.0 & 92.3 & 89.6 & 49.4  & 0.318 $\pm$ 0.011 & 16.76 & 17.77 \\
r05\_t60 & 20.0 & 64.4 & 62.9 & 85.9  & 0.250 $\pm$ 0.009 & 16.97 & 17.98 \\
r15\_t0  & 7.0  & 33.8 & 25.7 & 125.9 & 0.154 $\pm$ 0.004 & 17.44 & 18.44 \\
r10\_t0  & 5.0  & 26.1 & 18.6 & 127.9 & 0.150 $\pm$ 0.004 & 17.69 & 18.69 \\
r05\_t0  & 3.0  & 14.6 & 9.9  & 131.7 & 0.171 $\pm$ 0.005 & 18.29 & 19.29 \\
\hline
\end{tabular}
\end{table*}

MHONGOOSE provides six distinct resolution levels, constructed from different weighted combinations of baselines, with the lowest resolutions providing an increase of up to a factor of 100 in \HI{} column density sensitivity relative to the highest resolution \citep{MHONGOOSE}. The well-known trade-off between angular resolution and sensitivity in principle allows for an increase in the radial extent of kinematic models by utilising the lower resolution levels in the outermost regions of the galaxy where \HI{} emission is weakest. This was already possible in earlier surveys -- for instance, THINGS\footnote{The \HI{} Nearby Galaxies Survey} on the Very Large Array delivered `robust' and `natural' weighted cubes with differing spatial resolutions and sensitivities \citep{Walter2008} -- but MeerKAT has stable noise properties over a superior dynamic range in spatial resolution.

Many studies have combined rotation curves derived from different physical tracers. A common approach combines H$\alpha$ and \HI{} rotation curves \citep[e.g.][]{Swaters2000, deBlok2001, Swaters2003}. H$\alpha$ often offers superior angular resolution and so is used to probe the central regions, while \HI{} extends the curve to significantly larger radii. Millimetre/sub-millimetre observations of molecular emission has also been used towards the centre, once again supplemented by \HI{} to extend the radial coverage \citep[e.g.][combining CO and \HI{} observations]{Sofue1996}. Similarly, \cite{Leung2021} provide an example of combining stellar (optical) and HI-derived rotation curves to provide tighter constraints on dark matter models. Whilst the examples above combine different physical tracers, an alternative method involves combining data derived from the same physical tracer, often with varying levels of spatial resolution and sensitivity. This enables us to obtain rotation curves that benefit from both high resolution in the central regions and high sensitivity at larger galactocentric radii where emission is weakest. Most relevant to our work, \cite{Namumba2025} conducted a kinematic study of two galaxies from the MHONGOOSE survey, and combined models fitted to separate resolution levels to increase the radial extent of their rotation curves by several kiloparsecs. However, this was achieved by essentially `gluing' the lower resolution rotation curve onto the end of the higher resolution curve, with little regard for the physical consistency of the combined model. This leads to discontinuities in the combined model (e.g. inclination and position angle profiles, see their fig.~6) at the boundary between the two resolution levels.

This work aims to address these shortcomings by combining observations at the multiple resolution levels provided by MHONGOOSE in a physically consistent manner, obtaining physical parameters which are consistent across all resolution levels and taking advantage of both the high resolution and high sensitivity data where each is best used. This paper is structured as follows. Section~\ref{Sec 2} outlines the data used and target selection. Section~\ref{Sec 3} details the software configuration used to fit kinematic models to each galaxy. Section~\ref{Sec 4} explains the relative weightings of resolution levels used as a function of radius, while Section~\ref{Sec 5} describes the method by which we combine the available resolution levels using the above weighting functions and presents the results of applying this methodology to both mock and real data. Finally, in Section~\ref{Sec 6} we discuss our main findings, and summarise our key conclusions in Section~\ref{Sec 7}.

\begin{figure*}
    \includegraphics[width=1.0\textwidth]{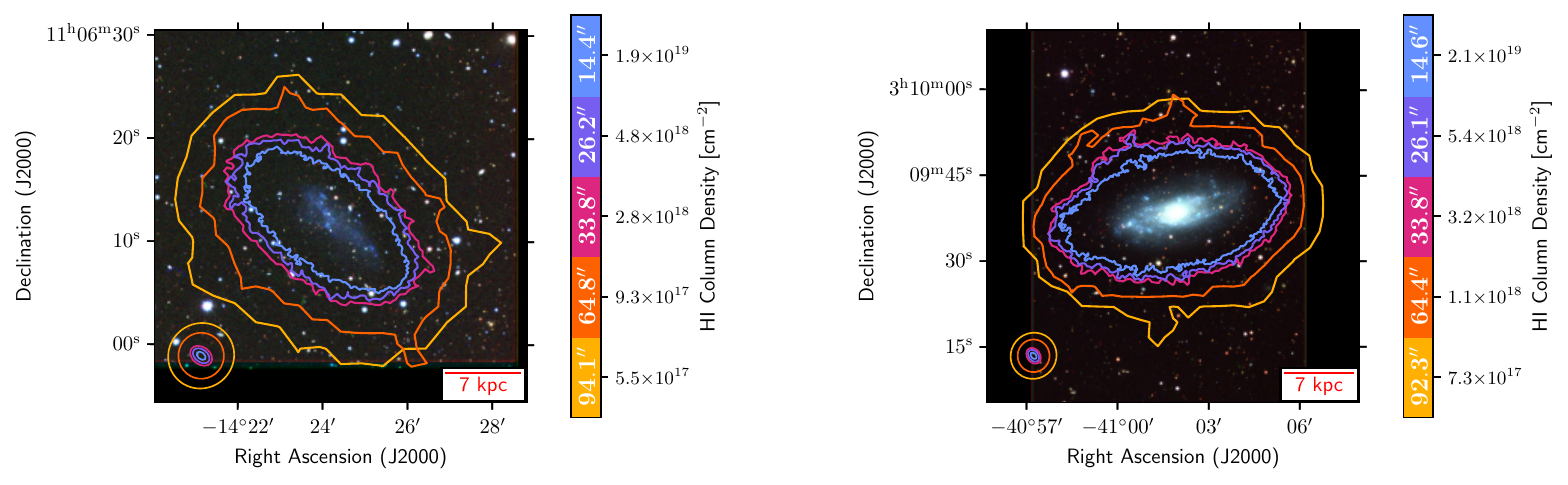}
    \caption{MHONGOOSE \HI{} $\mathrm{S}/\mathrm{N}=3$ contours at different resolution levels overlaid onto DECaLS \textit{grz} band images of J1106-14 (left) and J0309-41 (right). The contours are colour-coded according to the respective column density colour bars. These contours, in order from smallest to largest, are derived from the r05\_t0 (light blue), r10\_t0 (purple), r15\_t0 (pink), r05\_t60 (orange), and r10\_t90 (yellow) resolution levels. The values within the colour bars represent the beam major axis lengths for each resolution level. Concentric ellipses are included in the bottom left to represent the beam for each resolution, and a physical scale bar is included in the bottom right.}
    \label{fig: contour overlays}
\end{figure*}

\begin{figure*}
    \includegraphics[width=0.9\textwidth]{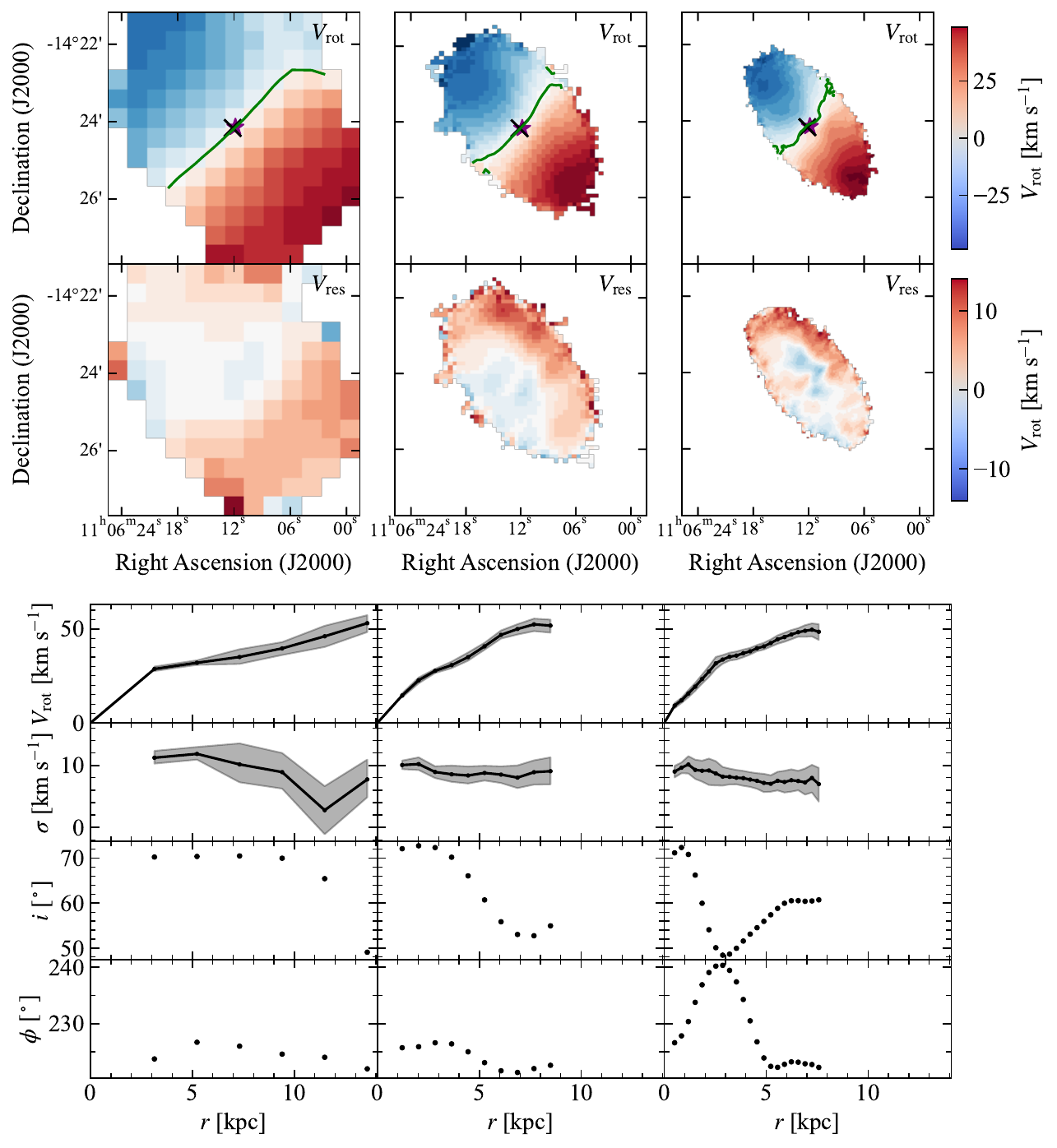}
    \caption{Best-fitting tilted ring models returned by \3DB{} for J1106-14 at the r10\_t90 (left column), r15\_t0 (middle column), and r05\_t0 (right column) resolution levels. The rotational (first row) and residual ($V_\mathrm{res} = V_\mathrm{rot} - V_\mathrm{model}$; second row) velocity maps are shown, along with profiles of the rotation speed (third row), velocity dispersion (fourth row), inclination (fifth row), and position angle (sixth row) as a function of radius. In the $V_\mathrm{rot}$ maps, the black cross and purple star indicate the dynamical centre positions as estimated by \3DB{} and given by the stellar light, respectively. The green line represents the zero isovelocity contour. All maps span the same angular region, and all profiles are displayed with the same radial limits. The shaded regions in the rotation speed and dispersion panels represent the uncertainties on these parameters returned by \3DB{}.}
    \label{fig: res models}
\end{figure*}

\begin{table*}
  \caption{Parameters used in our kinematical modelling. (1) HIPASS identification. (2) Right Ascension coordinate of dynamical centre position. (3) Declination coordinate of dynamical centre position. (4) Systemic velocity. (5) and (6) represent the initial inclinations (as in Table~\ref{table: Targets}) and position angles passed into 3D BAROLO.}
  \label{table: Modelling parameters}
  \begin{tabular}{ >{\centering\arraybackslash}m{2cm}  >{\centering\arraybackslash}m{3cm}  >
  {\centering\arraybackslash}m{3cm} >
  {\centering\arraybackslash}m{2cm} >
  {\centering\arraybackslash}m{1.5cm} >
  {\centering\arraybackslash}m{1.5cm}}
  \hline
    \makecell{Galaxy \\ \\ (1)} &
    \makecell{$\alpha_{\text{c}}$ (J2000) \\ ($^h\ ^m\ ^s$) \\ (2)} &
    \makecell{$\delta_{\text{c}}$ (J2000) \\ ($^\circ\ '\ ''$) \\ (3)} &
    \makecell{$v_{\text{sys}}$ \\ (km s$^{-1}$) \\ (4)} &
    \makecell{$i_\text{init}$ \\ ($^\circ$) \\ (5)} &
    \makecell{$\phi_\text{init}$ \\ ($^\circ$) \\ (6)} \\
    
    \hline
    
    J1106-14 & 11 06 12.1 & –14 24 10.0 & 1034 & 70 & 225 \\
    J0309-41 & 03 09 38.4 & -41 01 57.2 & 955.5 & 59 & 167 \\

    \hline
  \end{tabular}
\end{table*}

\section{Data and target selection}\label{Sec 2}

In this work, we focus on two galaxies from the MHONGOOSE survey with HIPASS\footnote{\HI{} Parkes All-Sky Survey} \citep{HIPASS} identifications J1106-14 and J0309-41. Key properties of these galaxies are summarized in Table~\ref{table: Targets}. These galaxies were chosen to be relatively simple to model: they are approximately symmetric both spatially and in velocity space. We also chose them to have different overall inclinations and stellar masses to provide two distinct test cases for our methodology.

For both of these galaxies we use five of the six available MHONGOOSE resolution levels, the beam sizes and \HI{} column density sensitivities for which are outlined in Table~\ref{table: resolution levels}. We omit the highest available resolution level since it does not reveal any additional features in the \HI{} intensity or velocity maps. Fig.~\ref{fig: contour overlays} shows the \HI{} contours derived from these five MHONGOOSE resolution cubes at a signal-to-noise of $3$ overlaid onto DECaLS\footnote{Dark Energy Camera Legacy Survey} \textit{grz} band images. We estimate the signal-to-noise ratio, $\mathrm{S}/\mathrm{N}$, as the ratio of the `signal' -- the sum of the flux density over a window of width $W_{20}$ \citep[taken from][]{MHONGOOSE} centred on the channel with maximum flux density in each pixel -- and the `noise' -- the root mean square scatter across pixels in a patch of the data cube well away from galactic emission and summed over the same spectral window as the signal. The figure shows both the increased radial extent of \HI{} relative to optical emission, and the relative increase in sensitivity at lower resolution levels. We note that the increasing radial extent may be mostly due to the increasing beam size; there may be little additional \HI{} at large radii \citep[see also][]{Namumba2025}.

\section{Kinematic modelling: \3DB{}}\label{Sec 3}

Obtaining well-constrained radial profiles from datacubes for key quantities, such as rotation curves, requires a detailed modelling approach. In this section, we outline the broadly conventional modelling procedure applied independently to each resolution level. The resulting models will be combined in subsequent sections.

\subsection{Tilted-ring modelling}
To obtain kinematic models of our target galaxies, we fit a 3D tilted-ring model to each datacube using the publicly available \3DB{} software\footnote{v1.7, git commit ID a47b6ca} \citep{DiTeodoro2015}. A 3D tilted-ring model is one in which the galactic \HI{} disk is broken down into a number of concentric rings, each with their own galactocentric radius ($r$), dynamical centre position ($\alpha_\mathrm{c}$, $\delta_\mathrm{c}$), systemic velocity ($v_\mathrm{sys}$), inclination ($i$), position angle ($\phi$), velocity dispersion ($\sigma$), and rotation speed ($v_{\mathrm{rot}}$). \3DB{} also estimates the \HI{} surface density profile, $\Sigma_\mathrm{HI}(r)$, by averaging the $21\,\mathrm{cm}$ line emission within each ring. We use \3DB{} over traditional 2D velocity field fitting algorithms, such as \textsc{diskfit} \citep{Sellwood2015} and \textsc{rotcur} \citep{Begeman1987}, because whilst 2D fitting algorithms are computationally inexpensive and produce reasonable results for high-resolution data, they are susceptible to the effects of beam smearing \citep[][this is particularly relevant since we use data with a very low spatial resolution]{Jozsa2007} and struggle to produce consistent results for galaxies with highly asymmetric profiles \citep{DiTeodoro2015}.

\begin{figure*}
    \includegraphics[width=1.0\textwidth]{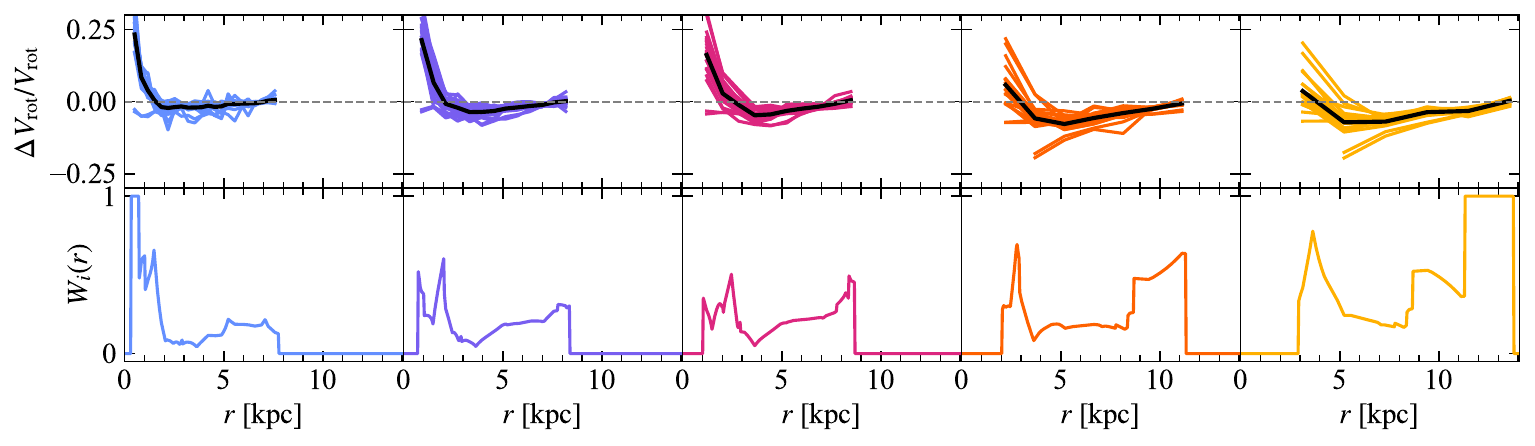}
    \caption{Calculation of radial weighting functions for J1106-14 across resolution levels: r05\_t0 (light blue), r10\_t0 (purple), r15\_t0 (pink), r05\_t60 (orange), and r10\_t90 (yellow) from left to right. The top row shows the fractional deviation of the rotation curve obtained by \3DB{} from the true curve for a range of mock galaxy geometries and kinematic configurations, with the average fractional residuals shown in black. The bottom row shows the resultant normalised weighting functions, derived from the inverse of the above residuals. These weighting functions represent the weighting \textit{per data point}, accounting for the radial sampling frequency of data at each resolution level.}
    \label{fig: weighting funcs}
\end{figure*}

\subsection{Model parameters}

\3DB{} identifies the best-fitting tilted-ring model by minimising the residuals between the data and model via its \textsc{3Dfit} task, returning parameters that have been corrected for inclination. It takes optional user-defined initial estimates for each model parameter (parameters for which the user does not provide an initial guess are estimated by the algorithm), as well as a set of parameters allowed to vary in the fitting process. We allow at most the rotational velocity, dispersion, inclination, and position angle of each ring in the model to vary; we fix the values of the centre position and systemic velocity. Since stars are more resilient against perturbation than gas, the peak of the stellar light provides a better tracer of the dynamical centre of a galaxy than the \HI{} gas \citep[e.g.][]{Oman2019}. Hence, we use DECaLS z-band ($\sim 0.9\mu$m) images to determine the intensity-weighted centre position of each galaxy via the distribution of older, redder stars. We then adjust the systemic velocity such that the dynamical centre position lies along the central zero isovelocity contour, as illustrated by the overlap of the black `X' marker and the green contour in the best-fitting models for J1106-14 in the top panel of Fig.~\ref{fig: res models}. The vertical thickness of the rings is fixed at $0\,\mathrm{pc}$ (razor-thin) unless explicitly stated otherwise. Whilst this is a poor approximation to the true thickness of galactic disks, \3DB{} is unable to self-consistently incorporate emission from the vertical components of adjacent rings -- a well-known limitation of tilted-ring modelling. Whilst \3DB{} can be configured to model only the approaching/receding side in isolation, we fit to both sides simultaneously due to the highly symmetric kinematic structure of the chosen galaxies, thus maximising the amount of information incorporated into the fitting process. The radial sampling is chosen to obtain $\sim$3 data points per beam, and the radial extent of our models are increased until a signal-to-noise threshold of $3$ is reached. Any rings that are flagged as unconverged fits by \3DB{} or are obviously erroneous are subsequently removed. The detailed configuration used for \3DB{} is given in Appendix~\ref{Appendix A}.

\3DB{} estimates the uncertainties on the rotation speed and velocity dispersion via Monte-Carlo sampling, by calculating the variation in these parameters that give rise to an up to 5~per~cent increase in residuals from a series of models constructed by changing the parameters with random Gaussian-distributed draws centred on the minimum in parameter space. Whilst this method is non-standard and computationally expensive, it returns uncertainties that are in agreement with other more standard methods \citep{DiTeodoro2015}, so we adopt them. It is worth noting that the uncertainties returned by \3DB{} do not influence our results, but are included for completeness.

\subsection{Two-stage fitting}

When fitting the inclination and position angle of each ring, to aid convergence we provide initial guesses as estimated from the \HI{} line-of-sight velocity map of each galaxy, allowing for deviations from these values of up to 40$^\circ$ in inclination and 20$^\circ$ in position angle. The initial estimates for inclination are derived from the \HI{} major-minor axis ratio, whilst initial estimates of the position angle are derived from the orientation of the major axis in the \HI{} moment-0 map. The major and minor axes are estimated by eye, and we ensure that these initial guesses agree well with the inclination values reported by \citet{MHONGOOSE} and the position angle values estimated by \3DB{} if no initial guess is provided. By default, \3DB{} implements a two-stage fitting algorithm when fitting these geometric parameters. Parameter regularisation occurs at the end of the first stage, where the inclination and position angle are `regularised' (interpolated) by a cubic B\'{e}zier (or other configurable) function. The dynamical parameters ($v_{\mathrm{rot}}$ and $\sigma$) are then fit again in the second stage using the regularised parameters from the first stage. Initial guesses are not provided for the rotation speed and velocity dispersion, as these have little impact on the final rotation curves and velocity dispersion profiles.

Fig.~\ref{fig: res models} shows the observed and residual velocity maps for J1106-14 at a low, medium, and high resolution level, alongside the radial rotation speed, velocity dispersion, inclination, and position angle profiles returned by the above two-stage fitting algorithm. The velocity maps highlight the trade-off between resolution levels: enhanced sensitivity and larger pixel size at lower resolutions, which together yield increased radial extent, but reduced spatial detail -- as demonstrated by the smoothness of the green isovelocity contour in the lower resolution map relative to those at higher resolutions. Additionally, the radial profiles in the bottom four rows of Fig.~\ref{fig: res models} highlight a clear inconsistency in the radial shape of the profiles obtained from different resolution levels; it is not simply a reduction in the frequency of data points in the lower resolution models. For example, the inclination and position angle profiles corresponding to the r05\_t0 model have sharp peaks/troughs at $\sim3\,\mathrm{kpc}$, whereas this feature is not present in the medium resolution level, and even less so in the lowest resolution level where the profiles are almost flat. Furthermore, the rotation curves at the lower resolution levels have a shallower increase in rotation speed than those at higher resolution levels, which we attribute to beam smearing.

This discussion demonstrates the motivation for combining multiple resolution levels. One may question whether the dramatic and perhaps unphysical variations in the geometry of the models at higher resolution levels is representative of the galaxy's true geometry, or whether the more smoothly varying profiles provided at lower resolution levels capture sufficient detail to pick up on smaller-scale fluctuations. By combining these models with appropriate weights, we aim to ensure that all available information is meaningfully incorporated, increasing confidence in the combined model, and importantly maintaining physical consistency across all resolution levels.

\begin{figure*}
    \includegraphics[width=0.95\textwidth]{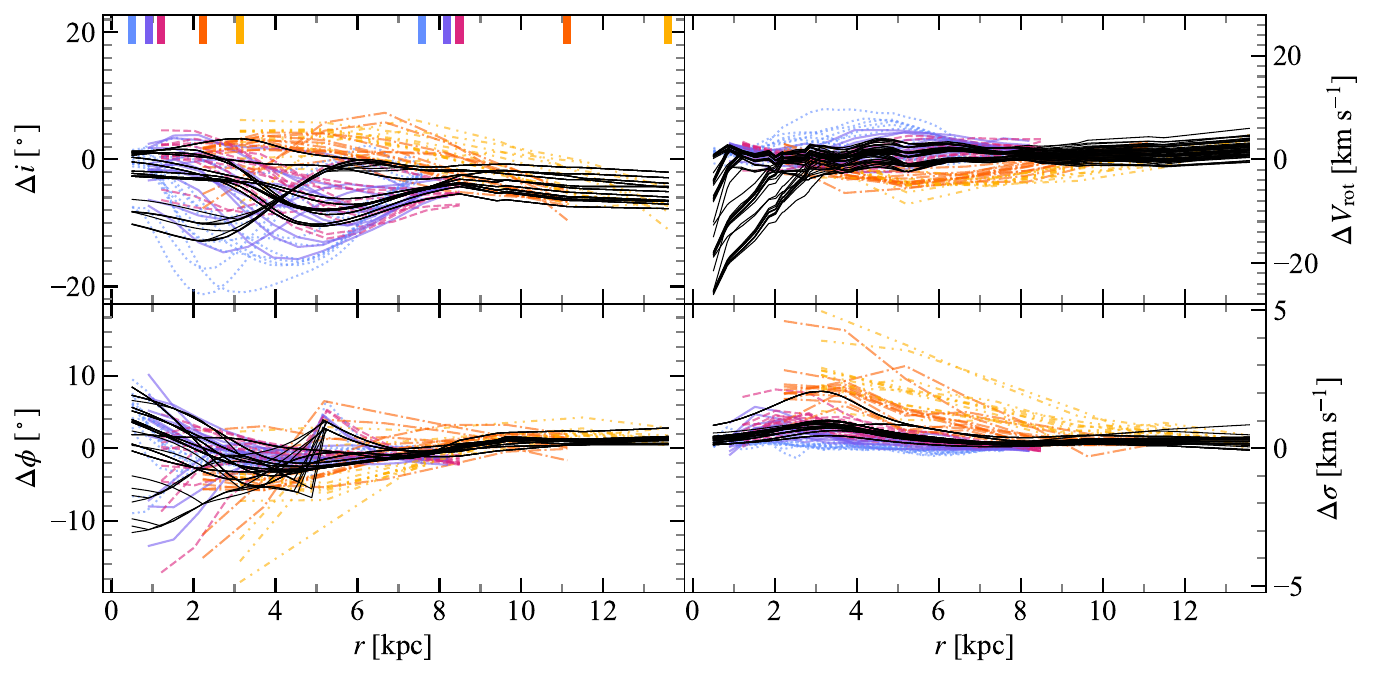}
    \caption{The application of our methodology to mock datacubes with known profiles. The coloured lines show the difference between the known profiles and those obtained from running \3DB{}'s two-stage fitting routine (with B\'{e}zier interpolation for the inclination and position angle profiles) on the 15 mock datacubes generated at each resolution level: r05\_t0 (dotted line; light blue), r10\_t0 (solid line; purple), r15\_t0 (dashed line; pink), r05\_t60 (dash-dot line; orange), and r10\_t90 (dash-double-dot line; yellow). The heavy black lines show the difference between the known profiles and those obtained via our combined smoothing method. The top left, bottom left, top right, and bottom right panels show the inclination, position angle, rotation speed, and velocity dispersion profiles, respectively. The vertical dashes in the top left panel mark the radii at which the data for each resolution level independently start and end.}
    \label{fig: apply to mock datacubes}
\end{figure*}

\section{Weighting functions}\label{Sec 4}

To combine kinematic models for multiple different resolution levels, we define relative weights for them as a function of radius. These weights are a local quantity - valid at a specific radius, and for a specific resolution level. To justify the form of these weighting functions, $W_i(r)$, we create a number of toy datacubes with known kinematic and geometric profiles using \3DB{}'s \textsc{galmod} task\footnote{Its \textsc{norm} parameter is set to \textsc{none}.}, and test how well the rotation curves can be recovered by the \textsc{3Dfit} routine. We create 15 such datacubes, each with different inclination, position angle, velocity dispersion, rotation velocity, and scale height radial profiles. A more detailed description of these profiles is given in Appendix~\ref{Appendix B}. We create a set of such datacubes for each target galaxy, using the corresponding observational datacubes as inputs (which set the beam size \& orientation, pixel size, etc.), and convert angular to physical scales using the distances in Table~\ref{table: Targets}. We then fit a tilted-ring model to each model datacube, once fitting only the rotation speed, and then again fitting both the rotation speed and velocity dispersion simultaneously. We choose to fix the geometric parameters to their true values in all cases as we find that \3DB{} struggles to recover the true inclination and position angle profiles regardless of the initial guesses provided, presumably due to the strong degeneracies present when these parameters are left free to vary.
 
At each resolution level, we calculate the fractional difference between the known rotation curve and that returned by the \textsc{3Dfit} routine, and then take the average of this over all 15 models with varying parameters.
\begin{equation}
\mathrm{mean}\left(\frac{V_\mathrm{rot,i}(r)-V_\mathrm{known,i}(r)}{V_\mathrm{known,i}(r)}\right) \equiv \mathrm{mean}\left(\frac{\Delta V_\mathrm{rot,i}(r)}{V_\mathrm{rot,i}(r)}\right).
\end{equation}
We then define the weighting function for a given resolution to be the inverse of the average fractional residual profile, with a small `softening parameter', $\epsilon$, included to prevent infinite values when the residuals approach zero (the index $i$ runs over the resolution levels from lowest to highest -- 0 to 4 in this case):
\begin{align}\label{equation: weighting func}
    \widetilde{W_i}(r) &= \frac{\Delta r_i/\Delta r_0}{\mathrm{mean}\left(\frac{\Delta V_\mathrm{rot,i}(r)}{V_\mathrm{rot,i}(r)}\right) + \epsilon};\\
    W_i(r)&=\frac{\widetilde{W_i}(r)}{\sum_i\widetilde{W_i}(r)}.
\end{align}
where $\widetilde{W_i}(r)$ is the non-normalised weighting function, $W_i(r)$ is normalised over all resolution levels, and both $\Delta r_i$ and $\Delta r_0$ are discussed below. The value of $\epsilon$ tunes the sensitivity of the weighting functions at small velocity residuals. In this work, we arbitrarily adopt $\epsilon = 10^{-2}$; the effect of changing this value is briefly explored in Appendix~\ref{Appendix C}. To account for the differing number of data points at each resolution level, the weighting function includes the ratio of the radial separation between data points at that resolution level ($\Delta r_i$) to the radial separation at the lowest resolution level ($\Delta r_0$). This ensures that higher resolution levels do not contribute more weight simply by virtue of being more finely sampled\footnote{Other weighting schemes could be considered here, such as the ratio of the ring areas, or the number of pixels in each ring. We choose the ratio of the ring spacings because it does not introduce an additional radial dependence in the weights. This is in the spirit of the \3DB{} geometric regularization approach where each ring is weighted equally, independent of its size.}. Finally, the weighting functions are interpolated onto a finer grid of radii, clipped to zero beyond the radial range of data at each resolution level, and then normalised such that they sum to unity at all radii.

Fig.~\ref{fig: weighting funcs} shows the velocity residuals and resulting weighting functions across all five resolution levels. The velocity residuals have a characteristic pattern at all resolutions, with the model rotation curves usually significantly overestimating the true values at small radii, before shooting down through the $\Delta V_\mathrm{rot}/V_\mathrm{rot} = 0$ line and converging towards the true value in their outermost regions. The overestimation of the rotation curve at small radii increases in radial extent at lower resolution levels, with the zero crossing of the average residuals occurring at the radius corresponding to approximately 1 beam width. In general, the lower resolutions carry more weight in the outermost regions of the galaxy as desired, with each $W_i(r)$ function exhibiting a characteristic `bump' between $2$ and $5\,\mathrm{kpc}$ where the residuals cross through zero.

\section{Combined geometric and kinematic models}\label{Sec 5}

Having defined weighting functions for each galaxy, we now integrate all five resolution levels into a unified model which captures both geometric structure and kinematic behaviour across scales. For each galaxy, we:
\begin{enumerate}[label=(\roman*), leftmargin=*, topsep=1pt]
    \item Fit a tilted-ring model to each resolution datacube using the \textsc{3Dfit} task, allowing $v_\mathrm{rot}$, $\sigma$, $i$, and $\phi$ to vary in each ring.
    \item Extract the inclination and position angles from the first-stage fitting (before regularization) of each resolution level and combine into one data vector.
    \item Perform a weighted Gaussian smoothing on the combined inclination and position angle datasets using the above weighting functions (which are constructed from a set of mocks matched to each galaxy's distance and size) and an adaptive Gaussian kernel width, $s(r)$, which is set to the distance to the $5^\mathrm{th}$ nearest data point. The kernel width is floored at $1\,\mathrm{kpc}$ for the inclination, position angle, and velocity dispersion profiles to prevent undersmoothing in the central regions (i.e. if the distance to the $5^\mathrm{th}$ nearest data point is less than $1\,\mathrm{kpc}$, $s(r)$ is set to $1\,\mathrm{kpc}$). It is instead floored to $0.25\,\mathrm{kpc}$ for the rotation curve due to it's steep incline. In all cases, the kernel width is capped at $3\,\mathrm{kpc}$ to prevent the smoothed profiles being strongly influenced by data from a large radial distance, particularly in the outermost regions where the data are most sparse. In particular, this prevents the rising portion of the rotation curve from exerting a strong influence on the smoothed curve at the largest radii.
    \item Re-fit a tilted-ring model to the datacube at each resolution with the geometric parameters ($i$ and $\phi$) fixed to their smoothed profiles as determined in the previous step.
    \item Extract the rotation curves and velocity dispersion profiles from each resolution level and combine into a single data vector.
    \item Reflect the rotation curve data in both the $r$- and $V_\mathrm{rot}$-axes to prevent oversmoothing at small radii and to enforce that the smoothed rotation curve passes through a rotation speed of zero at the dynamical centre. Reflect the velocity dispersion profile across $r=0$ -- no `vertical' reflection since $\sigma \ge 0$.
    \item Perform a weighted Gaussian smoothing on the reflected rotation curve and dispersion profile using the same weighting functions and kernel widths as above. Take the profiles at $r \ge 0$ as the smoothed, combined rotation curve and velocity dispersion profiles.
\end{enumerate}
Steps (i)-(iv) ensure that there is a consistent geometry across all resolution levels which, when combined with steps (v)-(vii), produces rotation curves and velocity dispersion profiles which utilise all of the available information. The choice to use the $5^\mathrm{th}$ nearest neighbour to define adaptive kernel width is somewhat arbitrary, but ensures that a sufficiently large number of data points are incorporated into the smoothing at large radii while not being so large that the profiles are oversmoothed in the central regions. A similar logic applies to the minimum ($1\,\mathrm{kpc}$ or $0.25\,\mathrm{kpc}$) and maximum ($3\,\mathrm{kpc}$) values of the kernel width. If the minimum value is too small, the profiles will be undersmoothed in the centre, but too large and we will lose sensitivity to features on kiloparsec scales. If the maximum value is too small, the outermost regions of the smoothed profile rely on very few data points; if it is too large, data at the edge of the model (around $r \approx 15\,\mathrm{kpc}$) can be overly influenced by intermediate rings (around $r \approx 10\,\mathrm{kpc}$). We further motivate these choices of values in Appendix~\ref{Appendix C}. Note that the minimum/maximum kernel widths we use are tailored to the two galaxies studied in this work, and are by no means general - further work would be required to generalise this methodology to all galaxies.

\begin{figure*}
    \includegraphics[width=0.83\textwidth]{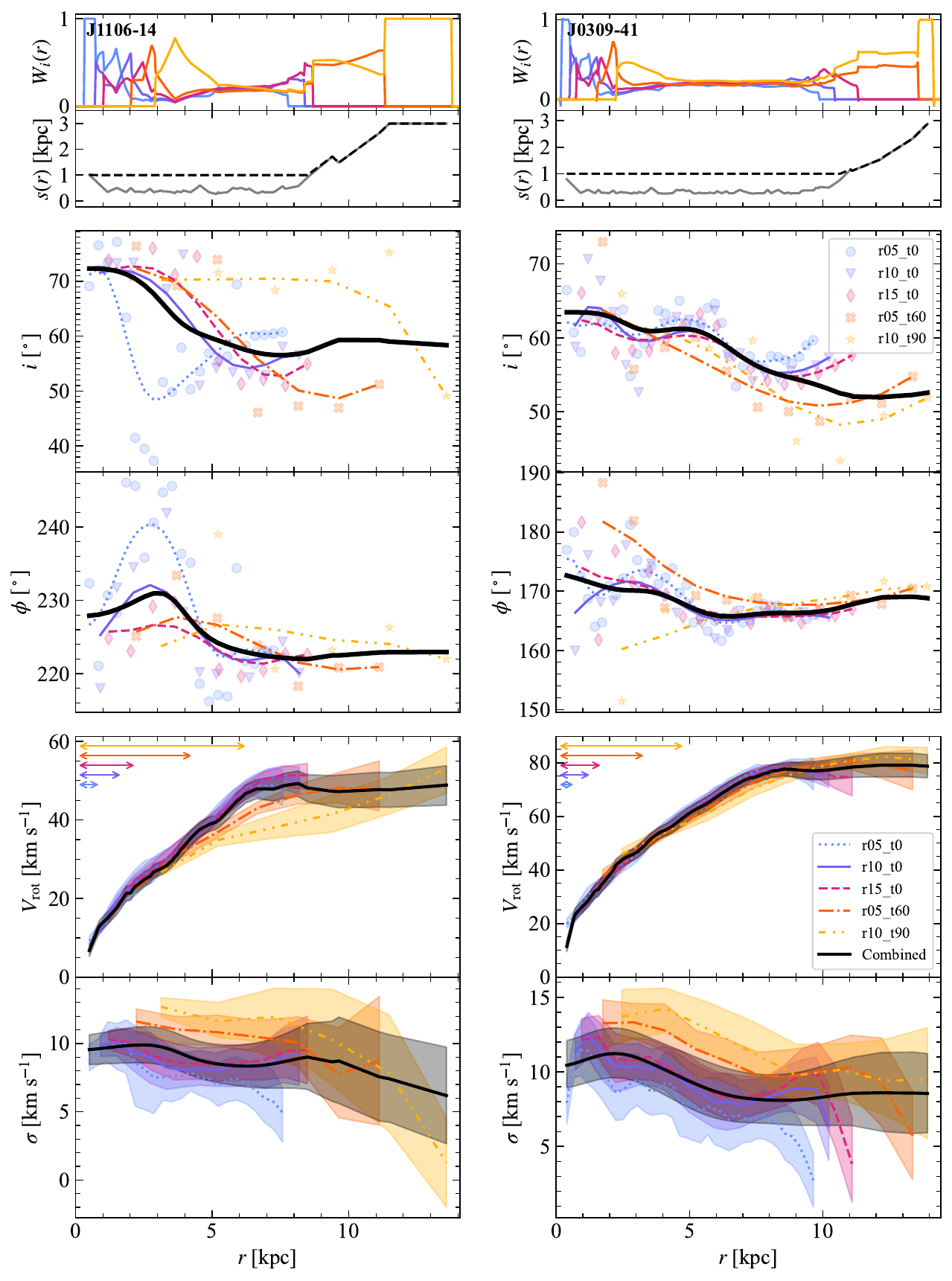}
    \caption{Demonstration of our multi-resolution approach for J1106–14 (left) and J0309–41 (right). The top row shows the \textit{per data point} weighting functions used in the smoothing, with the second row displaying the width of the Gaussian kernel as a function of radius. The dashed black $s(r)$ profile shows the kernel widths used to smooth the inclination, position angle, and velocity dispersion profiles, while the grey solid profile represents those used to smooth the combined rotation curve. The third and fourth rows present the inclination and position angle profiles obtained by independently fitting each resolution level (data points correspond to Stage 1 fits from \3DB{}; lines show profiles after Stage 2), alongside the combined, smoothed profiles shown by a heavy black line. The bottom two rows display the rotation curves and velocity dispersion profiles derived from each resolution level independently (coloured lines), using the combined inclination and position angle profiles. The smoothed curves, calculated using the same weighting functions, are shown in black, with shaded regions indicating the asymmetric uncertainty ranges. The double-sided arrows plotted in the $v_\mathrm{rot}$ panels represent the beam major axis lengths for each resolution level. The colour coding, linestyles, and marker styles used are as follows: r05\_t0 (dotted line; circle markers; light blue), r10\_t0 (solid line; triangle markers; purple), r15\_t0 (dashed line; diamond markers; pink), r05\_t60 (dash-dot line; 'X' markers; orange), and r10\_t90 (dash-double-dot line; star markers; yellow).}
    \label{fig: smoothing process}
\end{figure*}

\begin{figure*}
    \includegraphics[width=0.9\textwidth]{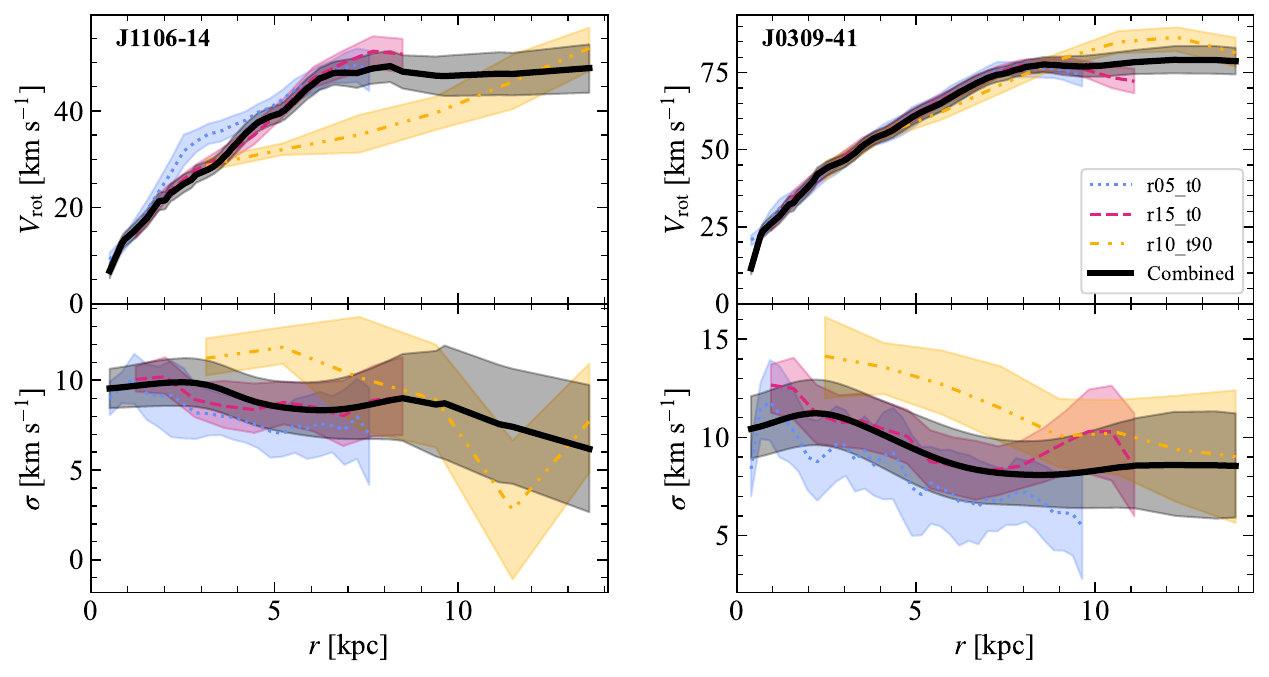}
    \caption{A comparison of the rotation curves (top row) and velocity dispersion profiles (bottom row) for J1106-14 (left) and J0309-41 (right) obtained via \3DB{}’s two stage fitting algorithm using a single resolution level only (coloured lines), and via our combined smoothing method (heavy black line). Only the lowest (dash-double-dot line; yellow), highest (dotted line; light blue), and intermediate (dashed line; pink) resolution levels are shown to improve clarity. Uncertainties estimated by \3DB{} are shown by the shaded regions.}
    \label{fig: final comparison}
\end{figure*}

\subsection{Mock datacubes}

Fig.~\ref{fig: apply to mock datacubes} shows the application of our smoothing process to the 15 mock datacubes generated previously. From the coloured lines, one can see that the models obtained from a single resolution level often struggle to accurately recover the known profiles. There are significant fluctuations in the geometric profiles of higher resolution models, whilst most resolutions fail to faithfully reproduce the kinematic profiles over their full radial extent. The combined geometric profile residuals (black lines) still exhibit a slight deviation from zero, which is to be expected since they are inherently constructed from the single-resolution models. However, these variations are significantly reduced, resulting in geometric profiles which generally lie much closer to the zero-residual line across all radii. This is similar to the combined kinematics. The combined rotation curves do not suffer from the severe overestimation in rotation speed seen at many resolution levels in their centre. Instead, they exhibit a significant underestimation in some cases. Both of these effects owe largely to our method of reflecting the rotation curves in both the radial and rotation speed axes before smoothing, as this ensures that the rotation curve passes through the origin at its centre as is physically required. The combined velocity dispersion profiles are quite accurate across their full extent, with deviations mostly within $1\,\mathrm{km\,s}^{-1}$. Furthermore, the central inaccuracies present in all individual resolution levels are completely eliminated. Overall, the combined profiles present a much more convincing model across all four parameters than any individual resolution level. This provides us with confidence that our combined smoothing method is able to provide more accurate results than using any single datacube in isolation across a range of galaxy geometries and kinematics.

\subsection{MHONGOOSE datacubes}

We now apply our methodology to real \HI{} observations. Fig.~\ref{fig: smoothing process} shows the application of our smoothing process to J1106-14 and J0309-41. As is evident in the second panel from the top of Fig.~\ref{fig: smoothing process} ($s(r)$), for both galaxies the kernel width for the $i$, $\phi$, and $\sigma$ profiles remains floored at $1\,\mathrm{kpc}$ across much of the radial extent of the combined model, but increases in the outermost regions where the sampling density of the data is much lower. The kernel width for the $V_\mathrm{rot}$ smoothing is smaller, as expected, but does not remain consistently floored at $0.25\,\mathrm{kpc}$. The third and fourth rows show the inclination and position angle profiles, respectively, derived from the initial fitting stage, the B\'{e}zier smoothing applied independently to each resolution level by \3DB{}, and the final combined profiles resulting from our smoothing procedure. The combined profiles capture the general trend of the unsmoothed data points, with a far smoother radial variation than any of the resolution levels analysed independently. The (probably unphysical) jumps in the higher resolution profiles are far less pronounced in the combined profiles, as evidenced by the position angle profiles of J1106-14 in the inner $5\,\mathrm{kpc}$. A significant `bump' of over 10$^\circ$ is seen in the r05\_t0 profile, and whilst this feature can still be seen in the combined profile, it is far less pronounced - an increase of only a few degrees from the central value. Similar arguments can be made for features in the other geometric panels in Fig.~\ref{fig: smoothing process}. This discussion demonstrates the advantage of utilising multiple resolution levels simultaneously: the combined profiles are still able to capture small-scale features that the lower resolutions independently cannot, but these fluctuations are significantly damped, producing much smoother and, in our view, physically reasonable geometries.

A similar smoothing technique is applied to the rotation curves and velocity dispersion profiles, as presented in the last two rows of  Fig.~\ref{fig: smoothing process}. The rotation curves returned by fitting to each resolution independently often struggle to converge to a rotation speed of zero at the dynamical centre, with this issue clearly mitigated by our strategy of including copies of the rotation curves reflected along both axes before smoothing. The combined profiles maintain the steady increases in rotation speed returned by the higher resolution models at intermediate radii, and do not adopt the likely incorrect shallower slope at intermediate radii ($5$ to $10\,\mathrm{kpc}$) at lower resolutions. However, the combined profiles still benefit from the additional sensitivity provided by lower resolutions in the outermost regions of the profiles, with an increase in radial extent of several kiloparsecs provided by the lowest two resolution levels.

\section{Discussion}\label{Sec 6}

Fig.~\ref{fig: final comparison} shows the final kinematic profiles returned by both our method of combining all resolution levels (in black, repeated from Fig.~\ref{fig: smoothing process}), and from running \3DB{} on a low, intermediate, and high resolution level independently via its two-stage fitting routine, as is common practice in kinematic studies of galaxies. The geometric profiles used in \3DB{}'s two-stage fitting routine for the latter are those shown by the dashed lines\footnote{The $V_\mathrm{rot}$ and $\sigma$ profiles are therefore not repeated from Fig.~\ref{fig: smoothing process} -- in that figure the combined profiles for the geometric parameters were used.} in Fig.~\ref{fig: smoothing process}. The combined kinematic profiles span a wider radial range than any single resolution level, allowing for an increase of $\sim50$~per~cent in radial coverage compared to the highest resolution level alone. The effects of beam smearing are most pronounced in the lowest resolution rotation curve, with a characteristic underestimation of the rotation curve slope and overestimation of the velocity dispersion within a scale of 1-2 beam sizes from the centre, as shown by the arrow markers in the $v_\mathrm{rot}$ panels of Fig.~\ref{fig: smoothing process}, eventually converging towards the correct value at the outermost radii. The combined rotation curves do not suffer from this issue, with a roughly linear increase in rotation speed at low radii (which well reflects the higher resolution data) and a subsequent plateau at larger radii (which converges towards the lowest resolution curve). This would, in principle, be possible to achieve by simply adding the very end of the lowest resolution curve onto that of the highest resolution, as in \cite{Namumba2025}, but this approach would lead to significant discontinuities in the geometric profiles; as illustrated by the stark difference between the blue and orange dashed lines in the $i$ and $\phi$ panels of Fig.~\ref{fig: smoothing process} -- such a model would not be self-consistent. Additionally, the single-resolution rotation curves often fail to converge to zero rotation speed at their centre, even at the highest resolution level. Our method of reflecting the $V_\mathrm{rot}$ profiles in both axes before smoothing ensures that the smoothed profiles satisfy $V_\mathrm{rot}(r=0)=0$ in all cases, and assist in preventing the erroneously large rotation speeds at the very centre of single resolution level models mentioned previously. Whilst it is possible to apply this technique to a single resolution, smoothing the rotation curve of a single resolution is conceptually more difficult to justify, whereas combining curves from multiple resolutions using appropriate weightings provides a clearer motivation for smoothing.

We have introduced a method for combining multiple resolution levels and demonstrated its benefits for kinematic and geometric profiles. However, it remains a demonstration rather than a fully optimised and generalised solution. The choice of weighting functions is the first point of discussion; while setting the weighting functions to the inverse of the $V_\mathrm{rot}$ residuals is well motivated, it could be improved upon. A better solution could incorporate the $i$, $\phi$, and $\sigma$ residuals, potentially using the average of all four to inform the weighting functions used. However, we are constrained by the current limitations of the \3DB{} software, as it struggles to accurately recover the true geometric (particularly inclination) and velocity dispersion profiles over the full radial range of the models, even from datacubes that it has generated itself. This often leads to large $i$, $\phi$, and $\sigma$ residuals. We explored this but found that this made it difficult to incorporate them without dominating the contribution from the rotation curve residuals, which we wished to emphasize. Future versions of the \3DB{} software are expected to improve on these limitations (E. di Teodoro, private communication), allowing for more accurate models and therefore a more thorough calculation of the weighting functions used. One may also note the jagged nature of the weighting functions used in Fig.~\ref{fig: weighting funcs}. While smoothing the average residual profile before taking the inverse would reduce these discontinuities, it would add an additional layer of complexity to our method and is unlikely to make a discernible difference to the resultant profiles. 

Another consideration is the chosen value of the softening parameter, $\epsilon = 10^{-2}$. Smaller values of $\epsilon$ make the weighting functions far more sensitive to smaller velocity residuals (potentially leading to the resolution with the smallest residuals at a given radius being given far too much weight relative to its counterparts), whereas larger values of $\epsilon$ reduce this sensitivity and lead to good and poor fits being treated more equally. We explore the effects of changing $\epsilon$ in more detail in Appendix~\ref{Appendix C}, but it is sufficient here to note that our chosen value strikes a reasonable balance between the considerations discussed above. There are also a number of parameters in our smoothing process which were chosen somewhat arbitrarily; namely the width of the Gaussian kernel (determined by the distance to the $N=5^\mathrm{th}$ nearest data point), and the minimum ($s_\mathrm{min} = 0.25\,\mathrm{kpc}$ for the rotation curve, or $1\,\mathrm{kpc}$ for all other parameters) and maximum ($s_\mathrm{max} = 3\,\mathrm{kpc}$) kernel widths. These values were briefly justified in Section~\ref{Sec 5}, but there is a range of values for each which yields similar results. This is also explored in Appendix~\ref{Appendix C}. 

Smoothing whilst preserving sufficient spatial resolution is an important trade-off in this methodology. Excessive smoothing erases small-scale structure, essentially eliminating the benefit of observations with high spatial resolution. On the other hand, insufficient smoothing results in profiles with sharp jumps and discontinuities. One may ask what the use of having observations with a spatial resolution of $\sim10''$ ($\sim0.67\,\mathrm{kpc}$ at a distance of $13.9\,\mathrm{Mpc}$) is if the kernel used for smoothing has a width much greater than this. From inspecting the geometric profiles in Fig.~\ref{fig: smoothing process}, our smoothed profiles do still preserve features from the highest resolution models, although they are less exaggerated. As previously noted, examples of this include the distinctive peak seen at $\sim 3\,\mathrm{kpc}$ in the combined position angle profile for J1106-14 in Fig.~\ref{fig: smoothing process} and the oscillations in the inclination profile of J0309-41. Both of these features are present (and more pronounced) in the higher resolution models, but their persistence in the combined profile shows that we are not completely erasing structure captured by the higher resolution data despite our smoothing procedure. Equally, we do not identify unphysical variations in our combined profiles, signifying that our chosen kernel width is sufficiently large. Our smoothing process also does not introduce additional freedom in the modelling process: it is loosely equivalent to the choice of regularization function in the conventional \3DB{} workflow (B\'{e}zier by default, but constant, linear and low-order polynomial functions are also often appropriate choices). Overall, we conclude that our combined profiles strike a reasonable balance between smoothing and preserving small-scale spatial features.

\section{Conclusions}\label{Sec 7}

In this work, we introduced a novel method of combining multiple \HI{} observations from the MHONGOOSE survey with different spatial resolutions while ensuring physical consistency across the combined model.

We derived a set of radial weighting functions by fitting tilted-ring models to datacubes with known geometric and kinematic profiles using the \3DB{} software at each resolution level. The models fail to recover the true profiles at small radii, often severely overestimating within the inner $\sim\mathrm{beam}$, followed by a zero crossing and a slow convergence toward zero residuals at the outer edge of the model. We only used $V_\mathrm{rot}$ residuals to calculate our weighting functions due to the inherent limitations of \3DB{} in fitting other parameters accurately, but there is scope in future work to incorporate $i$, $\phi$, and $\sigma$ residuals into this process using further iterations of the \3DB{} software.

Using these weighting functions, we combined observations of two MHONGOOSE galaxies at five resolution levels using adaptive Gaussian kernel smoothing. This method produces smoother, more physically plausible geometric profiles than any individual resolution level, with the sharp variations in $i$ and $\phi$ returned by the higher resolution models still present, but damped. The combined rotation curves and velocity dispersion profiles spanned a wider radial range than any individual resolution level, taking advantage of the high resolution data in the centre of the galaxy and the increased sensitivity of the lower resolution data in the outskirts. They suffer much less from both the erratic nature of the higher resolution data and effects such as beam smearing at lower resolutions. Our initial demonstration is promising, however the various parameters in both our determination of the weighting functions (i.e. the softening parameter, $\epsilon$) and our smoothing method (i.e. the number of nearest data points used to determine the kernel width, $N$, and the minimum \& maximum values of the kernel width, $s_\mathrm{min}$ \& $s_\mathrm{max}$) could be optimised further, perhaps informed by additional simulated datacubes that capture more of the complex features that arise in real galaxies.

\section*{Acknowledgements}

The authors would like to thank E.~de~Blok for helpful discussions on MHONGOOSE, and both E.~di~Teodoro and P.~Mancera-Pi\~{n}a for insightful advice on our use of the \3DB{} software.

KAO acknowledges support by the Royal Society through Dorothy Hodgkin Fellowship DHR/R1/231105, which also funded part of BRM's contribution. MS acknowledges the support by the UK Science and Technology Council (STFC) grant ST/X001075/1. This research has made use of NASA's Astrophysics Data System.

For the purpose of open access, the author has applied a Creative Commons Attribution (CC BY) licence to any Author Accepted Manuscript version arising from this submission.

Author contributions: BRM led analysis, created all figures, and drafted the manuscript, partially in fulfillment of requirements for an MPhys degree. KAO and MS designed and co-supervised the project and contributed to preparing the manuscript.

\section*{Data availability}

All \HI{} observations used in this work are available from the MHONGOOSE public data archive (\url{https://mhongoose.astron.nl/Public-Data-Release.html}). The optical g-, r-, and z-band images used to construct Fig.~\ref{fig: contour overlays} are available from the DECaLS image survey public archive (\url{https://www.legacysurvey.org/}).

\bibliographystyle{mnras}
\bibliography{references}

\appendix

\section{\3DB{} configuration}\label{Appendix A}

In Table~\ref{table: 3Dfit param file} we outline the configuration of \3DB{} used to fit tilted-ring models to datacubes via the \textsc{3Dfit} task. Only parameters that affect the result of the modelling are shown (i.e. file path definitions and diagnostic output parameters are omitted). All key parameters are shown, and any additional parameters which are not included are left at their default values, as given in the \3DB{} documentation\footnote{\url{http://editeodoro.github.io/Bbarolo/documentation/}}. This set of parameters was used to fit a tilted-ring model to each resolution level independently, from which the unsmoothed inclination and position angle profiles from the first stage fitting could be extracted. When the rotation speed and velocity dispersion were then fit with the geometric profiles fixed, `\textsc{free}' was set to `\textsc{vrot} \textsc{vdisp}', `\textsc{twostage}' was set to `\textsc{false}', and files containing the smoothed inclination and position angle profiles were passed into the `\textsc{inc}' and `\textsc{pa}' keywords - all other parameters remained unchanged. No initial guesses were provided for the rotation speed or velocity dispersion as the code is able to estimate these accurately without the need for user-defined initial values.

In Table~\ref{table: GALMOD param file}, we detail the configuration of \3DB{} used to generate the mock datacubes used to calculate our weighting functions via the \textsc{galmod} task. Once again, only parameters which affect the result of the modelling are shown, with any additional parameters left at their default values. Many optional parameters are the same as those used in the \textsc{3Dfit} task, but there are some subtle differences.

\begin{table*}
  \caption{Typical \textsc{3Dfit} parameter file for a single resolution level, omitting parameters that have no impact on the result of the calculation.}
  \label{table: 3Dfit param file}
  \begin{tabular}{ >{\centering\arraybackslash}m{3cm}  >{\centering\arraybackslash}m{3cm}  >
  {\centering\arraybackslash}m{2cm}  >
  {\arraybackslash}m{8cm} 
  }
  \hline
    \makecell{Keyword} &
    \makecell{Value} &
    \makecell{Units} &
    \makecell{Description} \\
    
    \hline
    
    3DFIT & TRUE & -- & Enable 3D fitting of datacube. \\
    NRADII & \textit{varies} & -- & Number of rings to fit. \\
    RADSEP & \textit{varies} & arcsec & Radial separation of rings. \\
    FREE & VROT VDISP INC PA & -- & Free fitting parameters. \\
    INC & as in Table~\ref{table: Modelling parameters} & deg & Initial inclination guess. \\
    PA & as in Table~\ref{table: Modelling parameters} & deg & Initial position angle guess. \\
    Z0 & \textit{varies} & arcsec & Disk scale-height (we assume a scale-height of 100pc, which corresponds to different angular scales at the differing distances of the two galaxies investigated). \\
    VSYS & as in Table~\ref{table: Modelling parameters} & km s$^{-1}$ & Systemic velocity. \\
    VRAD & 0 & km s$^{-1}$ & Radial velocity of rings. \\
    XPOS & as in Table~\ref{table: Modelling parameters} & deg & Dynamical centre position (R.A. coordinate). \\
    YPOS & as in Table~\ref{table: Modelling parameters} & deg & Dynamical centre position (Dec coordinate). \\
    DENS & 1 & atoms cm$^{-2}$ & Global column density of gas (unused when a normalisation is set). \\
    SIDE & B & -- & Kinematical side of the galaxy to fit to (B=both). \\
    NORM & AZIM & -- & Use azimuthal normalisation of rings. \\
    ADRIFT & FALSE & -- & Whether to perform asymmetric drift correction. \\
    VELDEF & RADIO & -- & Velocity definition to convert frequency/wavelength axis into velocity. \\
    FLAGERRORS & TRUE & -- & Estimate uncertainties via a Monte-Carlo approach. \\
    TWOSTAGE & TRUE & -- & Enables second stage fitting after parameter regularisation. \\
    REGTYPE & BEZIER & -- & Type of regularisation to use for second stage fitting (we choose regularisation via a Bezier function). \\
    DELTAINC & 40 & deg & Constrains inclination value to remain between [INC-DELTAINC, INC+DELTAINC]. \\
    DELTAPA & 20 & deg & Same as above but for position angle. \\
    MASK & SEARCH & -- & The source finding is run and the largest detection used to determine the mask. \\
    BLANKCUT & 2.5 & -- & Signal-to-noise threshold for mask construction. \\
    LTYPE & $1$ & -- & Layer type along z is Gaussian.\\
    FTYPE & $2$ & -- & Minimization function is $|{\rm model}-{\rm observed}|$.\\
    WFUNC & $2$ & -- & Azimuthal weighting function is $\cos^2\theta$.\\
    STARTRAD & $1$ & -- & Avoid the fit of the first ring \citep[see][]{Iorio2017}.\\
    \hline
  \end{tabular}
\end{table*}

\begin{table*}
  \caption{Typical \textsc{galmod} parameter file used in this work, omitting parameters that have no impact on the result of the calculation.}
  \label{table: GALMOD param file}
  \begin{tabular}{ >{\centering\arraybackslash}m{3cm}  >{\centering\arraybackslash}m{3cm}  >
  {\centering\arraybackslash}m{2cm}  >
  {\arraybackslash}m{8cm} 
  }
  \hline
    \makecell{Keyword} &
    \makecell{Value} &
    \makecell{Units} &
    \makecell{Description} \\
    
    \hline
    
    GALMOD & TRUE & -- & Enable 3D disk modelling. \\
    NRADII & \textit{varies} & -- & Number of rings to include. \\
    RADSEP & \textit{varies} & arcsec & Radial separation of rings. \\
    VROT & as outlined in Appendix~\ref{Appendix B} & km s$^{-1}$ & Rotation speed profile. \\
    VDISP & as outlined in Appendix~\ref{Appendix B} & km s$^{-1}$ & Velocity dispersion profile. \\
    INC & as outlined in Appendix~\ref{Appendix B} & deg & Inclination profile. \\
    PA & as outlined in Appendix~\ref{Appendix B} & deg & Position angle profile. \\
    Z0 & \textit{varies} & arcsec & Disk scale-height (we assume a scale-height of 100pc, which corresponds to different angular scales at the differing distances of the two galaxies investigated). \\
    VSYS & as in Table~\ref{table: Modelling parameters} & km s$^{-1}$ & Systemic velocity. \\
    VRAD & 0 & km s$^{-1}$ & Radial velocity of rings. \\
    XPOS & as in Table~\ref{table: Modelling parameters} & deg & Dynamical centre position (R.A. coordinate). \\
    YPOS & as in Table~\ref{table: Modelling parameters} & deg & Dynamical centre position (Dec coordinate). \\
    DENS & 1 & atoms cm$^{-2}$ & Global column density of gas. \\
    SIDE & B & -- & Kinematical side of the galaxy to fit to (B=both). \\
    NORM & NONE & -- & Use no normalisation. \\
    VELDEF & RADIO & -- & Velocity definition to convert frequency/wavelength axis into velocity. \\
    MASK & SEARCH & -- & The source finding is run and the largest detection used to determine the mask. \\
    LTYPE & $1$ & -- & Layer type along z is Gaussian.\\
    \hline
  \end{tabular}
\end{table*}

\section{Mock datacubes}\label{Appendix B}

\begin{figure*}
    \includegraphics[width=0.7\textwidth]{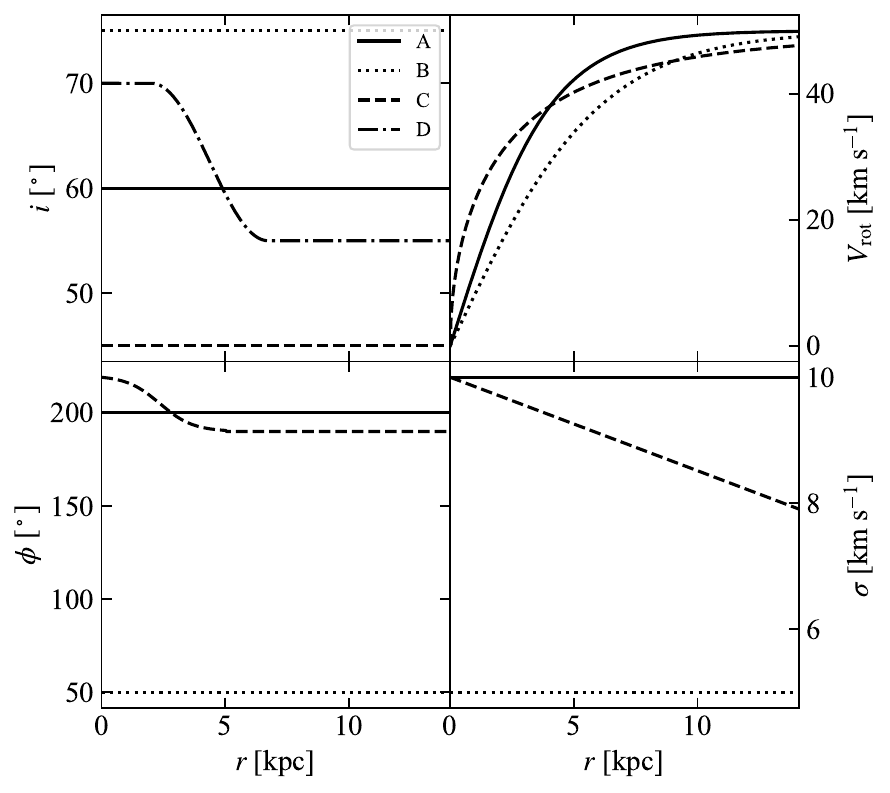}
    \caption{The different inclination (top left), position angle (bottom left), rotation speed (top right), and velocity dispersion (bottom right) profiles used to construct our mock datacubes for J1106-14. Those for J0309-41 have the same structure, but different physical radii on the x-axis. For each parameter, the different profiles are given differing linestyles, and are labelled from 'A' to 'D', as in the legend.}
    \label{fig: AppB}
\end{figure*}

In this appendix, we outline the profiles used to construct each of the 15 mock datacubes used in our work. Fig.~\ref{fig: AppB} shows each of the inclination, position angle, rotation speed, and velocity dispersion profiles used. Table~\ref{table: AppB} outlines the combination of profiles used for each mock datacube. The mock datacubes are constructed without the addition of noise -- in practice if noise was added we would then proceed to mask down to a high S/N subvolume of the mock cubes. Prior experience with \3DB{} \citep{Oman2019,Roper2023} has shown that this an inconsequential exercise, so we simply do not add any noise.

\begin{table*}
  \caption{Combination of profiles used to construct each of the 15 mock datacubes used in this work. All mock datacubes were constructed with a scale-height of $100\,\mathrm{pc}$ unless stated otherwise.}
  \label{table: AppB}
  \begin{tabular}{ >{\centering\arraybackslash}m{3cm}  >{\centering\arraybackslash}m{2cm}  >
  {\centering\arraybackslash}m{2cm}  >
  {\centering\arraybackslash}m{2cm}  >
  {\centering\arraybackslash}m{2cm}  >
  {\centering\arraybackslash}m{4cm}
  }
  \hline
    \makecell{Datacube Number} &
    \makecell{$i(r)$ profile} &
    \makecell{$\phi(r)$ profile} &
    \makecell{$V_\mathrm{rot}(r)$ profile} &
    \makecell{$\sigma(r)$ profile} &
    \makecell{Other} \\
    
    \hline
    
    1 & A & A & A & A & -- \\
    2 & B & A & A & A & -- \\
    3 & C & A & A & A & -- \\
    4 & A & B & A & A & -- \\
    5 & A & A & A & B & -- \\
    6 & D & A & A & A & -- \\
    7 & A & C & A & A & -- \\
    8 & D & C & A & A & -- \\
    9 & A & A & A & A & Zero scale-height \\
    10 & A & A & B & A & -- \\
    11 & A & A & C & A & -- \\
    12 & A & A & A & C & -- \\
    13 & D & C & A & A & -- \\
    14 & D & C & C & A & -- \\
    15 & D & C & C & C & Zero scale-height \\
    \hline
  \end{tabular}
\end{table*}

\section{Alternative weighting function and smoothing parameters}\label{Appendix C}

In this appendix, we briefly discuss and investigate how changing parameters used in both our determination of the weighting functions and our smoothing process affect the final results. The default set of parameters used are $N=5$, $\epsilon = 10^{-2}$, $s_\mathrm{min} = 1\,\mathrm{kpc}$ (or $0.25\,\mathrm{kpc}$ for $V_\mathrm{rot}$), and $s_\mathrm{max} = 3\,\mathrm{kpc}$.

\begin{figure*}
    \includegraphics[width=0.8\textwidth]{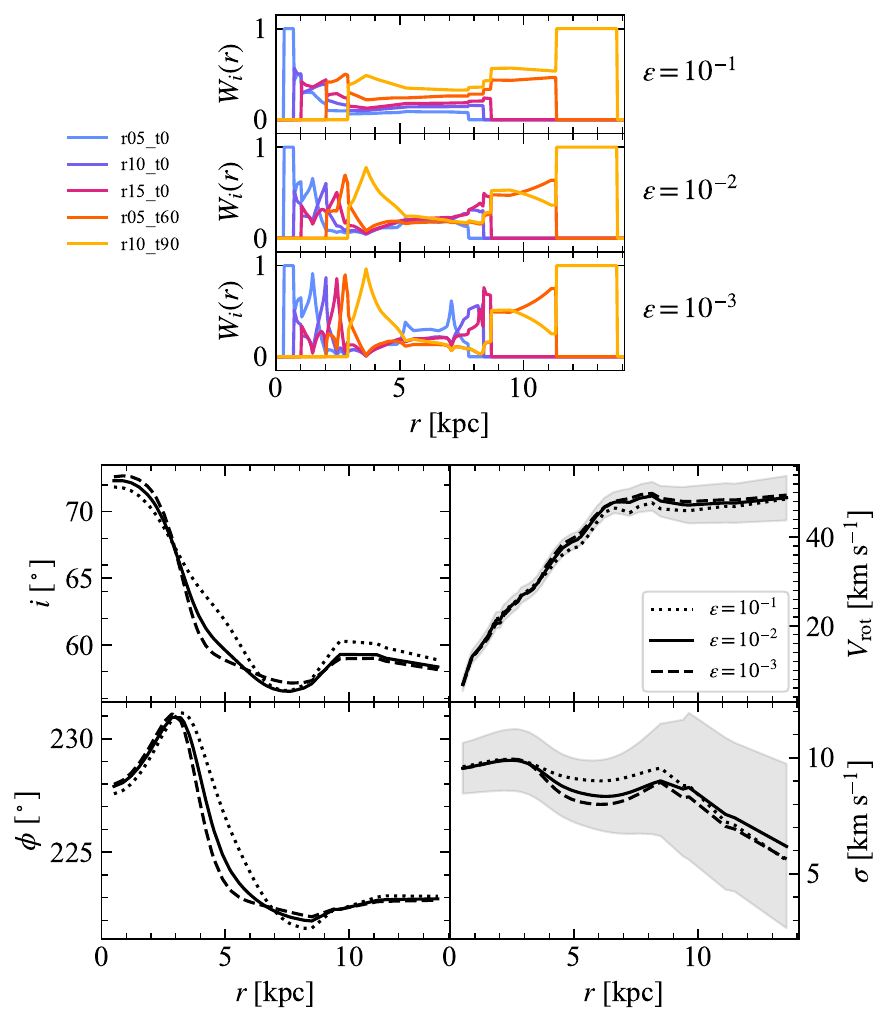}
    \caption{The weighting functions (top 3 panels) for three different values of the softening parameter, $\epsilon$, alongside the combined inclination (top left), position angle (bottom left), rotation speed (top right), and velocity dispersion (bottom right) profiles (bottom four plots) obtained via our smoothing process in each case. The profiles obtained using $\epsilon = 10^{-1}$, $10^{-2}$, and $10^{-3}$ are represented by the dotted, solid, and dashed lines, respectively.}
    \label{fig: AppC}
\end{figure*}

Fig.~\ref{fig: AppC} shows how both the weighting function, and the resultant geometric and kinematic profiles vary with three different values of $\epsilon$ spanning multiple orders of magnitude. As expected, smaller values of $\epsilon$ increase the sensitivity of the weighting functions to small residuals, thus accentuating sharp peaks. Smaller values of $\epsilon$ have the opposite effect. The inclination and position angle profiles in each case vary by $\sim$a few degrees, but the variation in the resultant kinematics lies well within the uncertainty range returned by \3DB{}.

The effects of changing the smoothing parameters (i.e. $N$, $s_\mathrm{min}$, and $s_\mathrm{max}$) are more intuitive. Decreasing the number of nearest neighbours, $N$, used to determine the kernel width leads to undersmoothed profiles with significant fluctuations on sub-kpc scales, whereas increasing $N$ far beyond $5$ produces oversmoothed profiles which completely eliminate the features picked up by higher resolution data, thus nullifying its use. Decreasing the minimum kernel width, $s_\mathrm{min}$, has a similar effect to decreasing N at radii where the density of data points is greatest. Conversely, increasing $s_\mathrm{min}$ parallels increasing $N$ since more data points will lie within one standard deviation ($1\sigma$) of the Gaussian kernel if $\sigma$ is larger. Increasing the maximum kernel width, $s_\mathrm{max}$, beyond $3\,\mathrm{kpc}$ only affects the outermost data point, with it being more strongly influenced by data at smaller radii. On the other hand, decreasing $s_\mathrm{max}$ below $3\,\mathrm{kpc}$ leads to an undersmoothing of data in the outermost regions of the galaxy since data is sparsest at large radii, resulting in sharp variations in the geometric and kinematic profiles towards their outer edge. The combination of $N$, $s_\mathrm{min}$, and $s_\mathrm{max}$ we have chosen in this work provides a reasonable balance between these issues. A fully general method to fix the smoothing parameters in particular is left to future work since it should likely depend on the physical and angular size of the galaxy, the angular resolutions of the observations and the strength of gradients (e.g. of the rotation curve and inclination profile), and our exploration with two galaxies does not sufficiently cover the parameter space needed to inform a generalised method.

\section{Tabulated combined kinematic and geometric profiles}\label{Appendix D}

The combined $V_\mathrm{rot}$, $\sigma$, $i$, and $\phi$ profiles from our smoothing method (and as plotted by the heavy black lines in Figs.~\ref{fig: smoothing process} and \ref{fig: final comparison}) are tabulated in Table~\ref{table: J1106-14 tabulated profiles} for J1106-14 and Table~\ref{table: J0309-41 tabulated profiles} for J0309-41.

\clearpage
\LTcapwidth=\textwidth
\renewcommand{\arraystretch}{1.3}
\onecolumn

\begin{longtable}{
  >{\centering\arraybackslash}m{2cm}
  >{\centering\arraybackslash}m{3cm}
  >{\centering\arraybackslash}m{3cm}
  >{\centering\arraybackslash}m{2cm}
  >{\centering\arraybackslash}m{2cm}
}
\caption{Kinematic and geometric parameters derived from our combined smoothing method for J1106-14. (1) Ring radius. (2) Rotational velocity with asymmetric errors. (3) Velocity dispersion with asymmetric errors. (4) Inclination. (5) Position angle.}
\label{table: J1106-14 tabulated profiles} \\
\hline
\makecell{Radius \\ (kpc) \\ (1)} &
\makecell{$V_{\text{rot}}$ \\ (km s$^{-1}$) \\ (2)} &
\makecell{$\sigma$ \\ (km s$^{-1}$) \\ (3)} &
\makecell{$i$ \\ ($^\circ$) \\ (4)} &
\makecell{$\phi$ \\ ($^\circ$) \\ (5)} \\
\hline
\endfirsthead

\multicolumn{5}{l}{{\textbf{Table \thetable{}.} -- \textit{continued}}} \\
\hline
\makecell{Radius \\ (kpc) \\ (1)} &
\makecell{$V_{\text{rot}}$ \\ (km s$^{-1}$) \\ (2)} &
\makecell{$\sigma$ \\ (km s$^{-1}$) \\ (3)} &
\makecell{$i$ \\ ($^\circ$) \\ (4)} &
\makecell{$\phi$ \\ ($^\circ$) \\ (5)} \\
\hline
\endhead
0.51 & $6.8^{+1.4}_{-1.5}$ & $9.5^{+1.1}_{-1.1}$ & 72.3 & 227.9 \\
0.84 & $12.6^{+1.4}_{-1.5}$ & $9.6^{+1.1}_{-1.1}$ & 72.3 & 228.1 \\
0.91 & $13.3^{+1.5}_{-1.5}$ & $9.6^{+1.1}_{-1.1}$ & 72.3 & 228.1 \\
1.18 & $15.1^{+1.6}_{-1.6}$ & $9.7^{+1.2}_{-1.2}$ & 72.2 & 228.3 \\
1.21 & $15.4^{+1.6}_{-1.6}$ & $9.7^{+1.2}_{-1.2}$ & 72.1 & 228.4 \\
1.52 & $17.7^{+1.7}_{-1.8}$ & $9.8^{+1.2}_{-1.2}$ & 71.8 & 228.8 \\
1.85 & $21.3^{+1.8}_{-1.9}$ & $9.8^{+1.3}_{-1.3}$ & 71.3 & 229.3 \\
2.02 & $21.4^{+1.8}_{-1.8}$ & $9.9^{+1.3}_{-1.3}$ & 70.9 & 229.6 \\
2.12 & $22.6^{+1.8}_{-1.8}$ & $9.9^{+1.3}_{-1.3}$ & 70.6 & 229.8 \\
2.19 & $23.2^{+1.8}_{-1.8}$ & $9.9^{+1.3}_{-1.3}$ & 70.4 & 229.9 \\
2.22 & $23.2^{+1.8}_{-1.8}$ & $9.9^{+1.3}_{-1.3}$ & 70.3 & 230.0 \\
2.53 & $24.9^{+1.9}_{-1.8}$ & $9.9^{+1.3}_{-1.3}$ & 69.3 & 230.5 \\
2.73 & $25.7^{+1.9}_{-1.7}$ & $9.9^{+1.4}_{-1.3}$ & 68.3 & 230.8 \\
2.83 & $26.5^{+1.9}_{-1.7}$ & $9.9^{+1.4}_{-1.3}$ & 67.9 & 230.9 \\
2.86 & $26.8^{+1.9}_{-1.7}$ & $9.8^{+1.4}_{-1.3}$ & 67.7 & 230.9 \\
3.13 & $27.6^{+1.8}_{-1.6}$ & $9.8^{+1.4}_{-1.3}$ & 66.2 & 230.9 \\
3.20 & $27.9^{+1.8}_{-1.6}$ & $9.7^{+1.4}_{-1.3}$ & 65.9 & 230.9 \\
3.34 & $28.5^{+1.8}_{-1.6}$ & $9.6^{+1.4}_{-1.3}$ & 65.1 & 230.7 \\
3.54 & $29.7^{+1.8}_{-1.7}$ & $9.5^{+1.4}_{-1.3}$ & 64.0 & 230.2 \\
3.64 & $30.6^{+1.9}_{-1.8}$ & $9.4^{+1.4}_{-1.3}$ & 63.5 & 229.8 \\
3.71 & $31.0^{+1.9}_{-1.8}$ & $9.4^{+1.4}_{-1.3}$ & 63.2 & 229.6 \\
3.87 & $32.6^{+2.0}_{-1.9}$ & $9.2^{+1.4}_{-1.3}$ & 62.4 & 228.8 \\
3.94 & $33.1^{+2.0}_{-1.8}$ & $9.2^{+1.4}_{-1.3}$ & 62.2 & 228.5 \\
4.21 & $35.3^{+2.0}_{-1.9}$ & $9.0^{+1.4}_{-1.3}$ & 61.3 & 227.3 \\
4.45 & $37.0^{+2.0}_{-1.9}$ & $8.8^{+1.4}_{-1.3}$ & 60.7 & 226.3 \\
4.55 & $37.6^{+1.9}_{-1.9}$ & $8.8^{+1.4}_{-1.3}$ & 60.4 & 225.9 \\
4.89 & $38.8^{+1.9}_{-1.8}$ & $8.6^{+1.4}_{-1.3}$ & 59.8 & 224.9 \\
5.16 & $39.5^{+1.8}_{-1.8}$ & $8.5^{+1.4}_{-1.4}$ & 59.3 & 224.3 \\
5.19 & $39.7^{+1.8}_{-1.9}$ & $8.5^{+1.4}_{-1.4}$ & 59.3 & 224.2 \\
5.22 & $40.0^{+1.9}_{-1.9}$ & $8.5^{+1.4}_{-1.4}$ & 59.2 & 224.2 \\
5.26 & $40.0^{+1.9}_{-1.9}$ & $8.5^{+1.4}_{-1.4}$ & 59.2 & 224.1 \\
5.56 & $42.2^{+2.0}_{-2.1}$ & $8.4^{+1.5}_{-1.4}$ & 58.6 & 223.6 \\
5.76 & $43.7^{+2.0}_{-2.2}$ & $8.4^{+1.5}_{-1.4}$ & 58.3 & 223.4 \\
5.90 & $44.7^{+2.1}_{-2.3}$ & $8.4^{+1.5}_{-1.4}$ & 58.0 & 223.2 \\
6.07 & $45.7^{+2.1}_{-2.3}$ & $8.3^{+1.6}_{-1.5}$ & 57.8 & 223.1 \\
6.23 & $46.8^{+2.2}_{-2.4}$ & $8.3^{+1.6}_{-1.5}$ & 57.5 & 222.9 \\
6.37 & $47.1^{+2.3}_{-2.4}$ & $8.3^{+1.6}_{-1.5}$ & 57.3 & 222.8 \\
6.57 & $47.8^{+2.3}_{-2.5}$ & $8.3^{+1.7}_{-1.6}$ & 57.1 & 222.7 \\
6.67 & $47.9^{+2.3}_{-2.5}$ & $8.4^{+1.7}_{-1.6}$ & 56.9 & 222.6 \\
6.87 & $47.9^{+2.6}_{-2.7}$ & $8.4^{+1.8}_{-1.7}$ & 56.8 & 222.5 \\
6.91 & $48.0^{+2.6}_{-2.8}$ & $8.4^{+1.8}_{-1.7}$ & 56.7 & 222.4 \\
6.97 & $47.9^{+2.7}_{-2.8}$ & $8.4^{+1.8}_{-1.7}$ & 56.7 & 222.4 \\
7.24 & $47.8^{+3.1}_{-3.1}$ & $8.5^{+1.9}_{-1.8}$ & 56.6 & 222.3 \\
7.31 & $48.0^{+3.1}_{-3.1}$ & $8.5^{+1.9}_{-1.8}$ & 56.6 & 222.3 \\
7.58 & $48.6^{+3.0}_{-3.1}$ & $8.6^{+2.0}_{-1.9}$ & 56.5 & 222.2 \\
7.68 & $48.8^{+3.1}_{-3.2}$ & $8.7^{+2.0}_{-2.0}$ & 56.5 & 222.1 \\
8.15 & $49.3^{+3.2}_{-3.3}$ & $8.9^{+2.3}_{-2.2}$ & 56.8 & 222.0 \\
8.19 & $49.2^{+3.2}_{-3.3}$ & $8.9^{+2.3}_{-2.2}$ & 56.8 & 222.0 \\
8.49 & $48.2^{+3.4}_{-3.5}$ & $9.0^{+2.5}_{-2.4}$ & 57.1 & 222.0 \\
9.40 & $47.3^{+4.0}_{-3.9}$ & $8.6^{+3.0}_{-2.9}$ & 58.9 & 222.5 \\
9.64 & $47.2^{+4.2}_{-4.1}$ & $8.7^{+3.2}_{-3.1}$ & 59.3 & 222.5 \\
11.12 & $47.7^{+4.4}_{-4.5}$ & $7.6^{+3.3}_{-3.2}$ & 59.3 & 222.9 \\
11.49 & $47.7^{+4.4}_{-4.4}$ & $7.4^{+3.2}_{-3.2}$ & 59.0 & 222.9 \\
13.58 & $48.9^{+4.9}_{-5.1}$ & $6.2^{+3.6}_{-3.5}$ & 58.4 & 222.9 \\
\hline
\end{longtable}

\begin{longtable}{
  >{\centering\arraybackslash}m{2cm}
  >{\centering\arraybackslash}m{3cm}
  >{\centering\arraybackslash}m{3cm}
  >{\centering\arraybackslash}m{2cm}
  >{\centering\arraybackslash}m{2cm}
}
\caption{Kinematic and geometric parameters derived from our combined smoothing method for J0309-41. (1) Ring radius. (2) Rotational velocity with asymmetric errors. (3) Velocity dispersion with asymmetric errors. (4) Inclination. (5) Position angle.}
\label{table: J0309-41 tabulated profiles} \\
\hline
\makecell{Radius \\ (kpc) \\ (1)} &
\makecell{$V_{\text{rot}}$ \\ (km s$^{-1}$) \\ (2)} &
\makecell{$\sigma$ \\ (km s$^{-1}$) \\ (3)} &
\makecell{$i$ \\ ($^\circ$) \\ (4)} &
\makecell{$\phi$ \\ ($^\circ$) \\ (5)} \\
\hline
\endfirsthead

\multicolumn{5}{l}{{\textbf{Table \thetable{}.} -- \textit{continued}}} \\
\hline
\makecell{Radius \\ (kpc) \\ (1)} &
\makecell{$V_{\text{rot}}$ \\ (km s$^{-1}$) \\ (2)} &
\makecell{$\sigma$ \\ (km s$^{-1}$) \\ (3)} &
\makecell{$i$ \\ ($^\circ$) \\ (4)} &
\makecell{$\phi$ \\ ($^\circ$) \\ (5)} \\
\hline
\endhead
0.40 & $11.5^{+2.2}_{-2.0}$ & $10.4^{+1.7}_{-1.5}$ & 63.5 & 172.7 \\
0.66 & $22.1^{+2.2}_{-2.0}$ & $10.6^{+1.7}_{-1.5}$ & 63.5 & 172.3 \\
0.71 & $23.4^{+2.3}_{-2.0}$ & $10.6^{+1.7}_{-1.5}$ & 63.5 & 172.3 \\
0.92 & $25.8^{+2.4}_{-2.2}$ & $10.7^{+1.7}_{-1.5}$ & 63.4 & 172.0 \\
0.95 & $26.0^{+2.4}_{-2.2}$ & $10.7^{+1.7}_{-1.6}$ & 63.4 & 171.9 \\
1.19 & $28.4^{+2.5}_{-2.4}$ & $10.9^{+1.7}_{-1.6}$ & 63.4 & 171.6 \\
1.45 & $32.2^{+2.6}_{-2.7}$ & $11.0^{+1.7}_{-1.6}$ & 63.3 & 171.2 \\
1.59 & $32.8^{+2.5}_{-2.6}$ & $11.1^{+1.7}_{-1.6}$ & 63.3 & 171.0 \\
1.66 & $34.1^{+2.5}_{-2.7}$ & $11.1^{+1.7}_{-1.6}$ & 63.2 & 170.9 \\
1.72 & $34.8^{+2.5}_{-2.7}$ & $11.1^{+1.7}_{-1.6}$ & 63.2 & 170.9 \\
1.74 & $35.0^{+2.5}_{-2.7}$ & $11.1^{+1.7}_{-1.6}$ & 63.1 & 170.8 \\
1.98 & $37.8^{+2.4}_{-2.7}$ & $11.2^{+1.7}_{-1.6}$ & 62.9 & 170.6 \\
2.14 & $39.9^{+2.3}_{-2.6}$ & $11.2^{+1.7}_{-1.6}$ & 62.7 & 170.4 \\
2.22 & $41.4^{+2.3}_{-2.7}$ & $11.2^{+1.7}_{-1.6}$ & 62.6 & 170.3 \\
2.25 & $41.9^{+2.3}_{-2.7}$ & $11.2^{+1.7}_{-1.6}$ & 62.5 & 170.3 \\
2.46 & $43.6^{+2.2}_{-2.6}$ & $11.2^{+1.7}_{-1.6}$ & 62.2 & 170.2 \\
2.51 & $44.1^{+2.2}_{-2.6}$ & $11.2^{+1.7}_{-1.6}$ & 62.1 & 170.2 \\
2.62 & $44.6^{+2.1}_{-2.5}$ & $11.2^{+1.7}_{-1.6}$ & 61.9 & 170.2 \\
2.77 & $45.4^{+2.1}_{-2.4}$ & $11.1^{+1.7}_{-1.6}$ & 61.6 & 170.1 \\
2.85 & $45.7^{+2.1}_{-2.3}$ & $11.0^{+1.7}_{-1.6}$ & 61.5 & 170.1 \\
2.91 & $46.0^{+2.1}_{-2.3}$ & $11.0^{+1.7}_{-1.6}$ & 61.4 & 170.1 \\
3.04 & $46.7^{+2.1}_{-2.2}$ & $10.9^{+1.7}_{-1.6}$ & 61.3 & 170.1 \\
3.09 & $47.2^{+2.1}_{-2.3}$ & $10.9^{+1.7}_{-1.6}$ & 61.2 & 170.1 \\
3.30 & $49.0^{+2.2}_{-2.3}$ & $10.8^{+1.7}_{-1.6}$ & 61.0 & 170.0 \\
3.49 & $51.0^{+2.2}_{-2.2}$ & $10.6^{+1.7}_{-1.6}$ & 60.9 & 170.0 \\
3.57 & $51.7^{+2.2}_{-2.2}$ & $10.6^{+1.7}_{-1.6}$ & 60.9 & 169.9 \\
3.83 & $53.7^{+2.1}_{-2.2}$ & $10.3^{+1.7}_{-1.7}$ & 60.9 & 169.6 \\
4.04 & $54.7^{+2.2}_{-2.2}$ & $10.2^{+1.7}_{-1.7}$ & 61.0 & 169.3 \\
4.07 & $54.8^{+2.2}_{-2.2}$ & $10.1^{+1.7}_{-1.7}$ & 61.0 & 169.3 \\
4.10 & $54.9^{+2.2}_{-2.3}$ & $10.1^{+1.7}_{-1.7}$ & 61.0 & 169.2 \\
4.12 & $55.0^{+2.2}_{-2.3}$ & $10.1^{+1.7}_{-1.7}$ & 61.1 & 169.2 \\
4.36 & $56.4^{+2.5}_{-2.5}$ & $9.9^{+1.7}_{-1.7}$ & 61.2 & 168.7 \\
4.52 & $57.6^{+2.5}_{-2.5}$ & $9.8^{+1.8}_{-1.7}$ & 61.2 & 168.4 \\
4.62 & $58.4^{+2.6}_{-2.6}$ & $9.7^{+1.8}_{-1.8}$ & 61.2 & 168.2 \\
4.76 & $59.4^{+2.6}_{-2.6}$ & $9.6^{+1.8}_{-1.8}$ & 61.2 & 167.9 \\
4.89 & $60.5^{+2.6}_{-2.6}$ & $9.5^{+1.8}_{-1.8}$ & 61.2 & 167.6 \\
4.99 & $61.1^{+2.5}_{-2.6}$ & $9.4^{+1.8}_{-1.8}$ & 61.2 & 167.4 \\
5.15 & $62.2^{+2.4}_{-2.6}$ & $9.2^{+1.8}_{-1.8}$ & 61.1 & 167.1 \\
5.23 & $62.6^{+2.4}_{-2.6}$ & $9.2^{+1.8}_{-1.8}$ & 61.0 & 166.9 \\
5.39 & $63.4^{+2.4}_{-2.7}$ & $9.1^{+1.8}_{-1.8}$ & 60.8 & 166.6 \\
5.42 & $63.5^{+2.4}_{-2.7}$ & $9.0^{+1.8}_{-1.8}$ & 60.8 & 166.6 \\
5.47 & $63.8^{+2.4}_{-2.7}$ & $9.0^{+1.8}_{-1.8}$ & 60.7 & 166.5 \\
5.68 & $64.9^{+2.5}_{-2.8}$ & $8.9^{+1.8}_{-1.8}$ & 60.4 & 166.2 \\
5.73 & $65.3^{+2.5}_{-2.7}$ & $8.8^{+1.8}_{-1.8}$ & 60.3 & 166.2 \\
5.95 & $66.8^{+2.6}_{-2.7}$ & $8.7^{+1.8}_{-1.8}$ & 59.9 & 165.9 \\
6.02 & $67.3^{+2.6}_{-2.7}$ & $8.7^{+1.8}_{-1.8}$ & 59.7 & 165.9 \\
6.21 & $68.6^{+2.6}_{-2.7}$ & $8.6^{+1.8}_{-1.8}$ & 59.3 & 165.8 \\
6.39 & $69.7^{+2.6}_{-2.6}$ & $8.5^{+1.8}_{-1.7}$ & 58.8 & 165.7 \\
6.42 & $69.9^{+2.7}_{-2.6}$ & $8.5^{+1.8}_{-1.7}$ & 58.8 & 165.7 \\
6.47 & $70.2^{+2.7}_{-2.6}$ & $8.4^{+1.8}_{-1.7}$ & 58.6 & 165.7 \\
6.66 & $71.4^{+2.7}_{-2.5}$ & $8.4^{+1.8}_{-1.7}$ & 58.2 & 165.8 \\
6.74 & $71.8^{+2.6}_{-2.4}$ & $8.3^{+1.8}_{-1.7}$ & 58.0 & 165.8 \\
6.90 & $72.7^{+2.5}_{-2.4}$ & $8.3^{+1.7}_{-1.7}$ & 57.6 & 165.8 \\
7.00 & $73.3^{+2.4}_{-2.3}$ & $8.3^{+1.7}_{-1.6}$ & 57.3 & 165.9 \\
7.27 & $74.3^{+2.3}_{-2.3}$ & $8.2^{+1.7}_{-1.6}$ & 56.7 & 166.0 \\
7.29 & $74.3^{+2.3}_{-2.3}$ & $8.2^{+1.7}_{-1.6}$ & 56.7 & 166.0 \\
7.37 & $74.4^{+2.3}_{-2.3}$ & $8.2^{+1.7}_{-1.6}$ & 56.5 & 166.1 \\
7.53 & $74.9^{+2.3}_{-2.3}$ & $8.2^{+1.7}_{-1.6}$ & 56.2 & 166.1 \\
7.56 & $75.0^{+2.3}_{-2.3}$ & $8.2^{+1.7}_{-1.6}$ & 56.2 & 166.2 \\
7.79 & $76.0^{+2.3}_{-2.2}$ & $8.1^{+1.7}_{-1.6}$ & 55.8 & 166.2 \\
7.85 & $76.0^{+2.3}_{-2.2}$ & $8.1^{+1.7}_{-1.6}$ & 55.7 & 166.3 \\
7.93 & $76.2^{+2.3}_{-2.3}$ & $8.1^{+1.7}_{-1.6}$ & 55.6 & 166.3 \\
8.06 & $76.8^{+2.3}_{-2.2}$ & $8.1^{+1.7}_{-1.6}$ & 55.4 & 166.3 \\
8.32 & $77.2^{+2.4}_{-2.4}$ & $8.1^{+1.7}_{-1.6}$ & 55.1 & 166.3 \\
8.56 & $77.6^{+2.4}_{-2.5}$ & $8.1^{+1.8}_{-1.6}$ & 54.8 & 166.3 \\
8.59 & $77.6^{+2.4}_{-2.6}$ & $8.1^{+1.8}_{-1.6}$ & 54.8 & 166.3 \\
8.72 & $77.6^{+2.6}_{-2.7}$ & $8.1^{+1.8}_{-1.6}$ & 54.7 & 166.3 \\
8.80 & $77.6^{+2.6}_{-2.7}$ & $8.1^{+1.8}_{-1.6}$ & 54.6 & 166.3 \\
8.85 & $77.5^{+2.7}_{-2.8}$ & $8.1^{+1.8}_{-1.7}$ & 54.5 & 166.3 \\
9.01 & $77.3^{+2.8}_{-2.9}$ & $8.1^{+1.8}_{-1.7}$ & 54.4 & 166.3 \\
9.12 & $77.2^{+2.9}_{-3.0}$ & $8.1^{+1.9}_{-1.7}$ & 54.3 & 166.3 \\
9.20 & $77.2^{+3.0}_{-3.1}$ & $8.1^{+1.9}_{-1.7}$ & 54.2 & 166.4 \\
9.27 & $77.1^{+3.1}_{-3.1}$ & $8.2^{+1.9}_{-1.7}$ & 54.2 & 166.4 \\
9.38 & $77.1^{+3.2}_{-3.2}$ & $8.2^{+1.9}_{-1.8}$ & 54.1 & 166.4 \\
9.64 & $77.0^{+3.5}_{-3.3}$ & $8.2^{+2.0}_{-1.8}$ & 53.8 & 166.5 \\
9.75 & $77.0^{+3.5}_{-3.3}$ & $8.2^{+2.0}_{-1.8}$ & 53.7 & 166.5 \\
9.83 & $77.0^{+3.6}_{-3.4}$ & $8.3^{+2.0}_{-1.9}$ & 53.6 & 166.5 \\
9.88 & $77.1^{+3.6}_{-3.4}$ & $8.3^{+2.0}_{-1.9}$ & 53.5 & 166.6 \\
10.23 & $77.3^{+3.8}_{-3.6}$ & $8.4^{+2.1}_{-2.0}$ & 53.1 & 166.8 \\
10.46 & $77.6^{+3.8}_{-3.7}$ & $8.4^{+2.2}_{-2.0}$ & 52.7 & 167.0 \\
10.65 & $77.8^{+3.8}_{-3.7}$ & $8.5^{+2.3}_{-2.1}$ & 52.5 & 167.2 \\
11.04 & $78.3^{+3.9}_{-3.8}$ & $8.5^{+2.4}_{-2.2}$ & 52.2 & 167.7 \\
11.10 & $78.4^{+3.9}_{-3.9}$ & $8.6^{+2.4}_{-2.2}$ & 52.1 & 167.8 \\
12.21 & $79.1^{+4.4}_{-4.4}$ & $8.6^{+2.7}_{-2.6}$ & 51.9 & 169.0 \\
12.29 & $79.1^{+4.4}_{-4.4}$ & $8.6^{+2.7}_{-2.6}$ & 52.0 & 169.0 \\
13.37 & $79.1^{+4.5}_{-4.4}$ & $8.6^{+2.8}_{-2.7}$ & 52.3 & 169.1 \\
13.92 & $78.7^{+4.3}_{-4.3}$ & $8.6^{+2.7}_{-2.6}$ & 52.6 & 168.8 \\
\hline
\end{longtable}

\bsp
\label{lastpage}
\end{document}